\documentclass[aps,reprint,pre]{revtex4-1}
\usepackage[utf8]{inputenc}
\usepackage{mathtools}

\usepackage{graphicx}% Include figure files
\usepackage{dcolumn}% Align table columns on decimal point
\usepackage{bm}% bold math
\usepackage{xcolor}
\usepackage{amsmath}
\usepackage{amssymb}
\usepackage{hyperref}

\begin{document}

\title{Comparison of Two Analytic Energy Balance Models Shows Stable Partial Ice Cover Possible for Any Obliquity}

%[0000-0001-6029-5216]
\author{Ekaterina Landgren} 
\email{ek672@cornell.edu}
\affiliation{Center for Applied Mathematics; Cornell University, 657 Rhodes Hall, Ithaca, NY 14853}

%[0000-0003-4325-8047]
\author{Alice Nadeau} 
\email{a.nadeau@cornell.edu}
\affiliation{Department of Mathematics, Cornell University, Ithaca, NY 14853}

\date{\today}

\begin{abstract}
In this study, we compare two analytic energy balance models with explicit dependence on obliquity to study the likelihood of different stable ice configurations. We compare the results of models with different methods of heat transport and different insolation distributions. We show that stable partial ice cover is possible for any obliquity, provided the insolation distribution is sufficiently accurate.  Additionally, we quantify the severity of the transition to the Snowball state as different model parameters are varied. In accordance with an earlier study, transitions to the Snowball state are more severe for higher values of the albedo contrast and energy transport across latitudes in both model;  however, we find that the Snowball transition is not equally likely across both models. This work is general enough to apply to any rapidly rotating planet and could be used to study the likelihood of Snowball transitions on planets within the habitable region of other stars.  
\end{abstract}

\maketitle

\section{Introduction}
\label{sec:intro}

The search for habitable exoplanets, perhaps hosting life, is one of the great endeavors of our time.  To aid the search, it is important to understand what planetary factors contribute to a planet's habitability for life as we know it.  In this direction, analytical and computational studies using climate models to investigate different planetary scenarios have been incredibly important for advancing our scientific understanding of exoplanet climates. Recent work has investigated how a planet's orbital parameters such as obliquity (e.g. \cite{armstrong2014effects,Rose2017}),  eccentricity (e.g. \cite{kane2020eccentricity}), or spin-orbit resonance (e.g. tidal locking \cite{checlair2017no,checlair2019ocean,checlair2019HZ}) may affect the planet's climate.  Other studies have looked at what role climatic elements such as sea ice drift (e.g. \cite{yue2020effect}), land albedo (e.g. \cite{rushby2020effect}), or ocean heat transport (e.g. \cite{checlair2019ocean}) may play in the long term habitability of a planet. This study adds to the rapidly growing body of literature on exoplanet climate by considering the role of albedo/temperature feedback in a zonally averaged energy balance model with explicit obliquity dependence. We compare the model output for two methods of modeling the energy transport across latitudes and find qualitatively similar results between them.

This study is an extension of the work in \cite{Rose2017}.  In that study, \cite{Rose2017} analyze an energy balance model with explicit obliquity dependence and heat transport modeled with a diffusion term.  They find an analytical solution for temperature as a function of sine of the latitude and obliquity based on the methods given in \cite{north1975analytical}.  Their work shows that partial stable ice cover is more likely for ice caps than for ice belts (the model scenario where ice advances from the equatorial region of the planet rather than the poles, e.g. as in Figure \ref{figure-1}) and identifies regions of parameter space where the Snowball catastrophe---where a small change in radiative forcing causes the planet to quickly become completely ice covered---may be avoided.

Here we compare the model used in \cite{Rose2017} in two ways: (1) to the same model but with a more accurate approximation to the insolation distribution and (2) to a similar energy balance model but with heat transport modeled with relaxation to the global mean average temperature.  Comparisons between energy balance models with diffusive and relaxation terms for heat transport have been conducted in the past for Earth (e.g. \cite{north1984,roe2010,walsh2016diffusive}); however, an analysis agnostic to the specificity of particular planets has yet to appear in the literature.

In \cite{Rose2017}, the incoming stellar radiation is approximated by a second degree polynomial in sine of latitude and cosine of the obliquity. A second degree polynomial is sufficient to capture the qualitative behavior of insolation distributions for planets with very low obliquities (between $0^\circ$ and $45^\circ$) and obliquities close to $90^\circ$ (between $65^\circ$ and $90^\circ$), but does not capture the qualitative distribution for planets between these ranges.\footnote{Note that annual average insolation distribution for rapidly rotating planets is symmetric about $90^\circ$ obliquity.  The distribution when obliquity is $\beta$ is the same as when it is $180-\beta$.  In the following article, we restrict our attention to obliquities between $0^\circ$ and $90^\circ$.}  To capture the behavior accurately, one needs at least a sixth degree polynomial in sine of the latitude and cosine of the obliquity \cite{Nadeau2017,dobrovolskis2020}.  
Here we show that the higher order approximation is a necessary one; we would not find that stable partial ice cover is possible at any obliquity without it. In particular, we find that stable partial ice cover is possible for any obliquity for both diffusive and relaxation to the mean models.

Further, we find that the mode of heat transport does not change the qualitative distribution of the likelihood of planets with stable partial ice cover. Here we show that stable partial ice cover is less likely for high-obliquity planets than for low-obliquity planets for both diffusive and relaxation to the mean models, agreeing with the results already shown in \cite{Rose2017}. We also show that low albedo contrast and low efficiency of heat transport favor stable partial ice cover. The qualitative similarities between the models is encouraging because although analytical solutions can be found in both cases, the analytical solutions for the relaxation to the mean model are more tractable as they can be written using elementary functions.

Following \cite{Rose2017}, we  consider the stability of equilibrium of the system as radiative forcing changes in the model. Physically this could be caused by atmospheric effects, such as changing greenhouse gases.  In particular we consider how different parameters in the model affect the Snowball catastrophe.  We consider the severity of the Snowball catastrophe bifurcation based on the distance between the bifurcation latitude and the Snowball state. We identify the regions in parameter space where more severe bifurcations occur since a severe Snowball catastrophe may have stronger signals in a planet's climate record and may be more likely to be observed. We show that the Snowball catastrophe is less severe in the relaxation to the mean model compared to the diffusion model.

Our paper is laid out as follows. In Section \ref{sec:annual-avg-EB}, we present the governing equations and a nondimensionalization of these equations to better quantify the effects of parameter changes on the behavior of the system. In Section \ref{sec:temp-eq}, we derive the equations that relate the latitude of the saddle node bifurcation to the corresponding parameter values. In Section \ref{sec:likelihood}, we calculate the relative likelihood of stable partial ice cover. In Section \ref{sec:partial-ice-snowball}, we classify the bifurcation into the Snowball state based on its severity. A discussion of the results follows in Section \ref{sec:discussion} and we conclude in Section \ref{sec:conclusion}. In the Appendix, we briefly analytically compare the equilibrium solutions of the diffusion and relaxation to the mean models.

\section{Governing Equations}
\label{sec:governing-eq}

\subsection{Annual Average Energy Balance}
\label{sec:annual-avg-EB}

We consider a one dimensional energy balance model of the form popularized by \cite{budyko1969effect}, \cite{sellers1969global} and \cite{north1975theory}.  These models describe the time evolution of temperature on a planet depending on the incoming and outgoing radiation and heat transfer across latitudes. We consider the model

  \begin{equation}
     R\frac{\partial T}{\partial t}=Qs(y,\beta)(1-\alpha(y, \eta))-(A+BT)+\mathcal F(T)
     \label{eq:dim_model}
\end{equation}   

where $T=T(y,t)$ is the annual zonally averaged temperature as a function of sine of latitude $y$ and time $t$ and $\mathcal F(T)$ will either be diffusion 
  \begin{align*}
    \mathcal F(T)=D\nabla^2T=D\frac{\partial}{\partial y}(1-y^2)\frac{\partial}{\partial y}T(y,t)
\end{align*}   
or relaxation to the global mean temperature
  \begin{align*}
    \mathcal F(T)=-C(T-\overline {T})=-C\left(T(y,t)-\int_{0}^1 T(\phi,t)d\phi\right).
\end{align*}   
As noted in \cite{roe2010}, both forms of $\mathcal{F}(T)$ are parameterizations of the divergence of the poleward heat flux.  
Scientific discussions for the terms in the relaxation to the mean model can be found in the literature, e.g. \cite{held1974simple,checlair2017no}.  Readers interested in a mathematical discussion should see for example \cite{Tung2007, widiasih2013dynamics, mcgehee2014, walsh2016diffusive,kaper2013mathematics}; and \cite{north1975theory}. 
In the above model, $y$ is the sine of latitude. Hemispheric symmetry is assumed here so that the latitude $y$ ranges from 0 (the equator) to 1 (the north pole). The ice line latitude (the boundary between the high and low albedo regions) is denoted by $\eta$.  The mean annual amount of incoming solar radiation (insolation) is represented by $Q$. The insolation distribution function $s(y,\beta)$ depends on the latitude and on the planetary axial tilt $\beta$. The co-albedo function $(1-\alpha(y,\eta))$ determines the proportion of incoming solar radiation absorbed by the planetary surface at each latitude. Outgoing radiation  $A+BT$ is in a linearized form.
The dimensional parameters and their values for the Earth can be found in many references such as \cite{Tung2007} or \cite{kaper2013mathematics}.  Nondimensionalized parameter values for Earth are given in Table \ref{table:nondimensional_params}.

\begin{figure}
  \centering
  \includegraphics[width=0.99\linewidth]{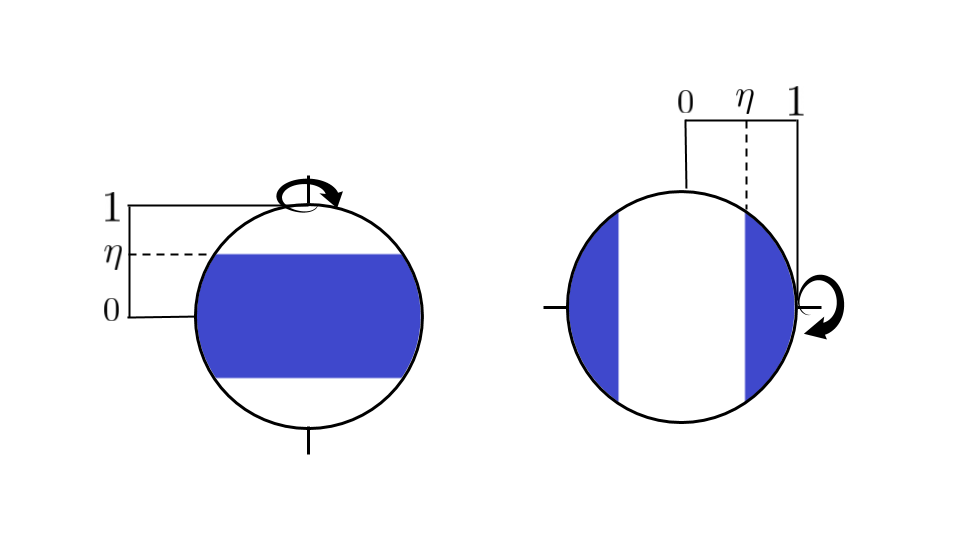}
  \caption{Planets at low obliquity (left) tend to exhibit ice caps, while planets at high obliquity (right) tend to exhibit ice belts. The ice line latitude is marked $\eta$.}
  \label{figure-1}
\end{figure}

The albedo function  $\alpha(y,\eta)$ is a piecewise constant function demarcating the regions of the planet  with high and low albedo.  Let $\alpha_p$ be the albedo polarward of the ice line, and $\alpha_e$ be the albedo equatorward of the ice line, then
  \begin{equation*}
\alpha(y,\eta)=
 \begin{cases} 
      \alpha_{e} & 0<y< \eta \\
      \frac{\alpha_e+\alpha_p}{2} & y=\eta \\
      \alpha_p & \eta<y< 1
   \end{cases}
\end{equation*}   
In the case of ice caps, the polar region has a higher albedo than the equatorial region with $\alpha_p>\alpha_e$.  For ice belts, the situation is reversed and $\alpha_p<\alpha_e$  (Figure \ref{figure-1}). In the following, we will denote high albedo with $\alpha_h$ (ice covered regions) and low albedo with $\alpha_l$ (non-ice covered regions).  In some energy balance models of Earth, the albedo values are set to  $\alpha_l=.32$ and $\alpha_h=.62$ with $\alpha_e=\alpha_l$ and $\alpha_p=\alpha_h$, (e.g. in \cite{widiasih2013dynamics}). Ice-line-dependent albedo is used in  \cite{mcgehee2014,widiasih2013dynamics,walsh2016diffusive,barry2014nonsmooth}. This is in contrast with the temperature-dependent albedo used in \cite{Rose2017,north1975theory}. Here we use the ice-dependent version because while  the two forms are equivalent in the diffusion model \cite{cahalan1979stability},  the temperature dependent version in the relaxation to the mean model can result in spurious temperature solutions \cite{walsh2015budyko}.

We model the annual average changes in the temperature profile by using zonally averaged mean annual insolation in equation  \eqref{eq:dim_model}. The mean annual insolation distribution $s(y,\beta)$ depends only on obliquity $\beta$ and latitude $y$. The insolation distribution is given by the integral (e.g. \cite{ward1974climatic,mcgehee2014})
  \begin{align*}
s(y,\beta)=\frac{2}{\pi^2}\int_{0}^{2\pi}\sqrt{1-(\sqrt{1-y^2}\sin\beta\cos\gamma-y\cos\beta)^2}d\gamma
\end{align*}   
where $\gamma$ is the planet's longitude and which is not amenable to analytical calculations in the models we consider.  Instead, \cite{Nadeau2021} show that we may write $s(y,\beta)$ as a Legendre series 
  \begin{align*}
    s(y,\beta)=\sum_{n=0}^\infty a_{2n}p_{2n}(\cos\beta)p_{2n}(y) 
\end{align*}   
where $p_{2n}$ is the Legendre polynomial of degree $2n$ and
  \begin{align*}
    a_{2n}=\frac{(-1)^n(4n+1)}{2^{2n-1}}\sum_{k=0}^n{2n \choose n-k}{2n+2k\choose 2k}{1/2\choose k+1}
\end{align*}   
using the standard notation
  \begin{align*}
{x\choose k}=\frac{x(x-1)\cdots(x-k+1)}{k!}.
\end{align*}   
Truncating the series for some fixed $N$ gives a degree $2N$ polynomial approximation $\sigma_{2N}$ to the true insolation distribution, where increasing $N$ decreases the root mean square error of the approximation \cite{Nadeau2021}.  \cite{Nadeau2017,Nadeau2021} show that the sixth degree approximation given by 
  \begin{equation}
\begin{aligned}
    \sigma_6(y,\beta)&=\sum_{n=0}^3 a_{2n}p_{2n}(\cos\beta)p_{2n}(y)\\
    &=1+a_2 p_2(\cos \beta)p_2 (y)+a_4 p_4(\cos \beta)p_4 (y)\\
    &\qquad+a_6 p_6(\cos \beta)p_6 (y),
    \label{eq:insolation}
\end{aligned}
\end{equation}   
where $a_2=-5/8$, $a_4=-9/64$, and $a_6=-65/1024$, and 
  \begin{equation*}
\begin{aligned}
    p_2(y)&=(3y^2-1)/2, \\
    p_4(y)&=(35y^4-30y^2+3)/8, \\
    p_6(y)&=(231y^6-315y^4+105y^2-5)/16
\end{aligned}
\end{equation*}   
is the smallest degree approximation which captures the qualitative shape of the distribution for all obliquities and approximates the true distribution to within 1.6\% for all obliquities. In particular, it is the smallest degree approximation which captures the characteristic `W' shape of insolation distributions for planets with obliquity between approximately $45^\circ$ and $65^\circ$ \cite{Nadeau2017}. \cite{dobrovolskis2020} shows that extending the approximation to degree 8 or 10 improves the approximation significantly at the poles for obliquities close to zero but only slightly at other latitudes or for higher obliquities. In \cite{Rose2017}, the degree two approximation is used ($a_4=0$ and $a_6=0$). Notice that when $\cos\beta=\sqrt{3}/3$ (i.e. when $\beta\approx54.74^\circ$), the degree two approximation is identically equal to 1 for all values of $y$.  Here we use a generic truncation $\sigma_{2N}$ in our analysis.  In Section \ref{sec:likelihood} we present results for the cases where $N=1$ and $N=3$.

The position of the ice lines, denoted by $\eta$, depends on the mean annual temperature of the ice line.  Ice--albedo feedback is incorporated by the dynamic ice line equation that is coupled with equation \eqref{eq:dim_model}. We use the following dynamic ice line equation, first formulated in \cite{widiasih2013dynamics} for ice caps on Earth as
  \begin{equation}
    \frac{d\eta}{dt}=\rho(T(\eta,t)-T_c).
    \label{eq:ice_line_feedback}
\end{equation}   
For ice belts, the right-hand side should be multiplied by negative one. The physical boundaries at the pole and the equator are built into the model, i.e. $\eta$ cannot be greater than 1 or less than 0. A mathematical treatment of the resulting nonsmooth system using a projection rule and a Filippov framework can be found in \cite{barry2014nonsmooth}, where the invariance of the physically possible region is shown.

An intuitive interpretation of Equation \eqref{eq:ice_line_feedback} notes that the mean annual temperature at the ice line is denoted by $T(\eta,t)$. The critical temperature $T_c$ is the highest temperature at which multiyear ice can be present. If the ice line temperature is above $T_c$, then the ice cover shrinks. If the temperature at the ice line is below $T_c$, the ice cover grows. The response constant $\rho$ controls the speed of the ice line response to a change in temperature.   We are interested in the equilibrium position of the ice line, which is obtained when the ice line temperature is exactly $T_c$.

We nondimensionalize the system using transformations analogous to those in \cite{Rose2017}, namely 
  \begin{equation*}
    \tau=\omega t=\frac{2 \pi t}{t_{\text{year}}}, \ T^*=\frac{A+BT(y)}{A+BT_c},
\end{equation*}   
where $\omega=2 \pi/t_{\text{year}}$ rescales time to be dimensionless, such that $\tau=1$ corresponds to one year.
The nondimensionalized temperature $T^*$ is proportional to temperature and outgoing longwave radiation. At the ice line at equilibrium given by equation \eqref{eq:dynamic_ice_line}, $T^*$ is always equal to 1, i.e. $T^*(\eta)=1$.

The nondimensionalized parameters are summarized in Table \ref{table:nondimensional_params}. The parameter transformations (which are the same for both diffusion and relaxation to the mean) are
  \begin{align*}
\gamma&=\frac{R\omega}{B},&q&=\frac{(1-\alpha_l)Q}{A+BT_{c}},\\
\overline{\alpha}&=1-\frac{1-\alpha_h}{1-\alpha_l}, &\zeta&=\cos(\beta), \qquad \text{and} \\
\lambda&=\frac{\rho (A+B T_c)}{B \omega}.
\end{align*}   
These parameters have the following physical interpretations:
\begin{itemize}
\item $\gamma$: Seasonal heat capacity of the system relative to the outgoing radiation over one year.
\item $q$: A measure of radiative forcing balance. It is directly proportional to the annual average incoming solar radiation and inversely proportional to the outgoing radiation at critical temperature $T_c$.
\item $\overline\alpha$: A measure of albedo contrast that changes the ice\textemdash albedo feedback. The minimum value of $\overline\alpha=0$ means that the high albedo regions, $\alpha_h$, and the low albedo regions, $\alpha_l$,  have the same albedo. The maximum value of $\overline\alpha=1$ is not necessarily the maximal albedo contrast, but is instead the range of states where $\alpha_h=1$ (the regions with high albedo are infinitely reflective and absorb no energy) and $\alpha_l\not=1$ (the regions with low albedo absorb some energy).

\item $\lambda$: A measure of the speed of ice line response to the changes in temperature.

\end{itemize}
The nondimensional parameter for heat transport in the diffusion model is
  \begin{align*}
    \delta=\frac{D}{B}
\end{align*}   
and it is a measure of the efficiency of heat transport across latitudes \cite{Rose2017,stone1978}. The nondimensional parameter for heat transport in the relaxation to the mean model has the same form
  \begin{align*}
    \mu=\frac{C}{B}
\end{align*}   
and similar interpretation. Note that despite similar forms, $\mu$ does not directly correspond to $\delta$. The discrepancy is caused by the fact that the diffusion coefficient in the diffusion model is not a linear scaling of the horizontal heat transfer coefficient $C$ in relaxation to the mean model. \cite{north1975analytical} comments that the divergence operator and the relaxation to the mean operator have the same effect on degree two polynomials when $\delta=\mu/6$. When $N=3$, the relationship between these parameters is more complicated. We consider the relationship between equilibrium solutions to the two models in more detail the Appendix. Throughout this work, we use $\delta=\mu/6$ when directly comparing the behavior of the two models for fixed values of heat transport parameter.

Rewriting the albedo function with the nondimensionalization for ice caps yields
  \begin{equation*}
\alpha^*(y,\eta)=
 \begin{cases} 
      1 & 0<y< \eta \\
      \frac{2-\overline\alpha}{2} & y=\eta \\
      1-\overline\alpha & \eta<y< 1
   \end{cases}.
\end{equation*}   
For belts, the positions of $1$ and $1-\alpha$ are swapped with $\alpha^*(y,\eta)=1$ for $y>\eta$ and $\alpha^*(y,\eta)=1-\overline{\alpha}$ for $y<\eta$.

%\begin{widetext}
\begin{table}
\centering
\caption{Nondimensional Parameters for \eqref{eq:nondim_model} and \eqref{eq:nondim_model_diff}.\footnote{The dimensional parameter values are the same as in \cite{Tung2007}, except $D$, which is chosen to be consistent with \cite{Rose2017} and tuned to reflect the conditions of modern day climate (e.g. \cite{north1975theory}).}}
\label{table:nondimensional_params}
\begin{tabular}{||c |c |p{27mm} |p{13mm} ||}
\hline
Parameter&  Definition & Brief description  & Value for Earth\\ [0.2ex] 
 \hline\hline
$\gamma$ & $\frac{R\omega}{B}$ & Seasonal heat capacity   & 6.13 \\
 \hline
$q$ & $\frac{(1-\alpha_l)Q}{A+BT_{c}}$ & Radiative forcing &  1.27\\
  \hline
$\overline\alpha$ & $1-\frac{1-\alpha_h}{1-\alpha_l}$  & Albedo contrast  & 0.44\\
   \hline
$\zeta$&   $\cos(\beta)$ & Cosine of obliquity  & 0.92 \\
 \hline
$\lambda$ & $\frac{\rho (A+B T_c)}{B \omega}$  & Ice line response  & varies \\
 [1ex] 
 \hline
$N$&    & $2N$ is the degree of the insolation approximation  & 1 \\
 \hline
$\mu$ & $\frac{C}{B}$ & Relaxation to mean efficiency of heat transport    & 1.6 \\
  \hline
$\delta$ & $\frac{D}{B}$ & Diffusion efficiency of heat transport   & 0.31 \\
  \hline
\end{tabular}
\end{table}
%\end{widetext}

The nondimensionalized version of the temperature model \eqref{eq:dim_model} when heat transport is modeled as relaxation to the annual average temperature is given by
  \begin{equation}
    \gamma \frac{\partial T^*}{\partial \tau}=q \sigma_{2N}(y,\zeta)\alpha^*(y,\eta)-T^*-\mu(T^*-\overline{T^*})
    \label{eq:nondim_model}
\end{equation}   
and when heat transport is modeled as diffusion it is given by
  \begin{equation}
    \gamma \frac{\partial T^*}{\partial \tau}=q \sigma_{2N}(y,\zeta)\alpha^*(y,\eta)-T^*+\delta\nabla^2T^*.
    \label{eq:nondim_model_diff}
\end{equation}   
Note that $T^*$ is a function of $y$, $\eta$ and $\tau$ but those dependencies are being suppressed in the above equations.  In the relaxation to the mean model, $\overline{T^*}$ depends on $\eta$ and $\tau$ (see Section \ref{sec:temp-eq} for more details). The nondimensionalized ice-line equation  for ice caps  is 
\begin{equation}
   \frac{\partial \eta}{\partial \tau}=\lambda(T^*(\eta,\tau)-1),
   \label{eq:dynamic_ice_line}
\end{equation}
where $T^*(\eta,\tau)$ is the temperature evaluated at the ice line $y=\eta$. Note that as with the dimensional equation \eqref{eq:ice_line_feedback}, the right-hand side must be multiplied by $-1$ for ice belts. In the following we do not explicitly consider dynamics of the ice line and so $\lambda$ plays no role in the following analysis.

\subsection{Analysis of Temperature Equilibria}
\label{sec:temp-eq}
We expect to observe planets that have existed for a long time, and therefore we expect in the simplest case to find the corresponding planets at temperature---ice equilibrium. We focus on the equilibria of the ice line $\eta$ as they undergo bifurcations in radiative forcing $q$ and the associated hysteresis loops in our non-smooth system. Previously, bifurcations in $A$ and $Q$ have been considered for Earth's range of parameter values (\cite{widiasih2013dynamics}) and bifurcations in $q$ have been studied in \cite{Rose2017}. We follow the framework developed in  \cite{Rose2017} and study bifurcations in $q$ by considering a illustrative sample of combinations of the physical parameters.  In Section\ref{sec:likelihood} we will also conduct a parameter sweep in the obliquity, albedo contrast, and efficiency of heat transport to determine likelihood of stable partial ice cover.

In order to compute the parameter values that correspond to the bifurcations, we first need to find an expression for the equilibrium temperature profile  $T^*(y,\tau)$. Below we compute the equilibrium temperature profile for the relaxation to the mean model. This derivation can be found in many places (e.g. \cite{mcgehee2014}), so we only summarize it here for convenience.  Following these calculations, we summarize the method to compute the equilibrium for the diffusion model which is shown in detail in \cite{north1975analytical} and also summarized in \cite{Rose2017}. We briefly compare these analytical results in the Appendix.

\subsubsection{Relaxation to the Mean Equilibrium Temperature}

To find $T_{\text{eq}}^*(y,\eta)$, the equilibrium temperature at each latitude, we set
 $\frac{\partial T^*}{\partial t}=0$ and first find $\overline{T_{\text{eq}}^*}(\eta)$, the global average equilibrium temperature, by averaging over the latitudinal range $0$ to $1$. Since $\int_0^1 T^* (y,\eta,\tau)dy=\overline{T^*}(\eta,\tau)$, the last term in \eqref{eq:nondim_model} cancels, and we can explicitly solve for $\overline{T_{\text{eq}}^*}(\eta)$, namely 
  \begin{equation*}
    \overline{T_{\text{eq}}^*}(\eta)=\int_0^1 q \sigma_{2N} (y,\zeta) (\alpha^*(y,\eta)) dy.
\end{equation*}   
Note that $T_{\text{eq}}(\eta)$ depends on $\eta$ since $\alpha^*(y,\eta)$ depends on $\eta$, while the dependence on $y$ is integrated out.

Since $\alpha^*(y,\eta)$ differs for ice belts and ice caps, $\overline{T_{\text{eq}}^*}(\eta)$ also differs. Both solutions are polynomials depending on obliquity and ice edge latitude. For ease of notation, we let 
  \begin{align*}
    \Sigma_{2N}(\eta)=\int_0^{\eta} \sigma_{2N}(y,\zeta)dy=\sum_{n=0}^Na_{2n}p_{2n}(\zeta)P_{2n}(\eta)%=1-\int_{\eta}^1 \sigma_{2N}(y,\zeta)dy,
\end{align*}   
where $P_i(y)=\int p_i (y) dy$ is the antiderivative of the Legendre polynomial $p_i(y)$.  Then the global average equilibrium temperature is
   \begin{align}
    \overline{T_{\text{eq}}^*}(\eta)&=\begin{cases}
    q\Big((1-\overline\alpha)+\overline\alpha \Sigma_{2N}(\eta)\Big),&\text{ice caps}\\
    q\Big(1-\overline\alpha \Sigma_{2N}(\eta)\Big), &\text{ice belts}
    \end{cases}.
    \label{eq:Teq}
\end{align}   
Note that $\overline{T_{\text{eq}}^*}(\eta)$ is proportional to the nondimensionalized radiative forcing $q$.  For fixed $\eta$, the more radiative forcing the planet receives, the warmer its mean equilibrium temperature.

Once $\overline{T_{\text{eq}}^*}(\eta)$ has been found, we may solve for $T_{\text{eq}}^*(y,\eta)$ in Equation \eqref{eq:nondim_model}  with $\frac{\partial T^*}{\partial t}=0$. Indeed, for $y\not=\eta$
  \[T_{\text{eq}}^*(y,\eta)=\frac{1}{1+\mu}\left(q\sigma_{2N}(y,\zeta)\alpha^*(y,\eta)+\mu\overline{T^*_{\text{eq}}}(\eta)\right)\]   
Due to the discontinuity in $\alpha^*(y,\eta)$, the temperature profile is discontinuous at the ice line. We define the value at the ice line to be the average of the left and right limits of $T^*_{\text{eq}}$ as $y$ approaches $\eta$, namely
  \begin{equation*}
    T_{\text{eq}}^*(\eta,\eta)=\frac{\lim_{y\rightarrow \eta^+}T_{\text{eq}}^*(y,\eta)+\lim_{y\rightarrow \eta^-}T_{\text{eq}}^*(y,\eta)}{2}.
\end{equation*}   
For ice caps the left and right hand limits are
  \begin{align}
    \lim_{y\rightarrow \eta^-}T_{\text{eq}}^*(y,\eta)&= \frac{ q \sigma_{2N}(\eta,\zeta)+ \mu \overline{T_{\text{eq}}^*}(\eta)}{1+\mu},
    \label{eq:warm-branch}\\
   \lim_{y\rightarrow \eta^+}T_{\text{eq}}^*(y,\eta)&= \frac{ q \sigma_{2N}(\eta,\zeta)(1-\overline\alpha)+ \mu \overline{T_{\text{eq}}^*}(\eta)}{1+\mu}
   \label{eq:cold-branch}
\end{align}   
and for ice belts $\lim_{y\rightarrow \eta^-}$ and $ \lim_{y\rightarrow \eta^+}$ are swapped. In both cases, the equilibrium temperature at the ice line is given by 
  \begin{equation*}
\begin{aligned}
    T_{\text{eq}}^*(\eta,\eta)&=\frac{ q \sigma_{2N}(\eta,\zeta)(2-\overline\alpha)+ 2\mu \overline{T_{\text{eq}}^*}(\eta)}{2(1+\mu)};
\end{aligned}
\end{equation*}   
however, the function for $\overline{T_{\text{eq}}^*}(\eta)$ differs for ice caps and ice belts (see Equation \eqref{eq:Teq}). Note that this definition of the temperature profile at the ice line coincides with the result derived using our definition of $\alpha^*(\eta,\eta)$.

Ice line equilibria occur when $T_{\text{eq}}^*(\eta,\eta)=1$. We consider the response of equilibria to changes in radiative forcing $q$.
The relaxation to the mean model allows us to solve exactly for the unique value of radiative forcing $q$ as a function of the other parameters, namely 
  
\begin{equation}
    q_{\eta}(\zeta,\overline\alpha,\mu)=\frac{2(1+\mu)}{\sigma_{2N}(\eta,\zeta)(2-\overline\alpha)+2\mu T_x(\eta,\zeta)},
    \label{eq:q_eta}
\end{equation}   
where 
  \begin{align*}
    T_x(\eta,\zeta)=
    \begin{cases}
    (1-\overline\alpha)+\overline\alpha \Sigma_{2N}(\eta,\zeta)&\text{ice caps}\\
    1-\overline\alpha \Sigma_{2N}(\eta,\zeta) &\text{ice belts}
    \end{cases}.
\end{align*}   
Note that $T_x(\eta,\zeta)=\overline{T_{\text{eq}}^*}(\eta)/q$. 

\begin{figure}
  \centering
  \includegraphics[width=0.45\linewidth]{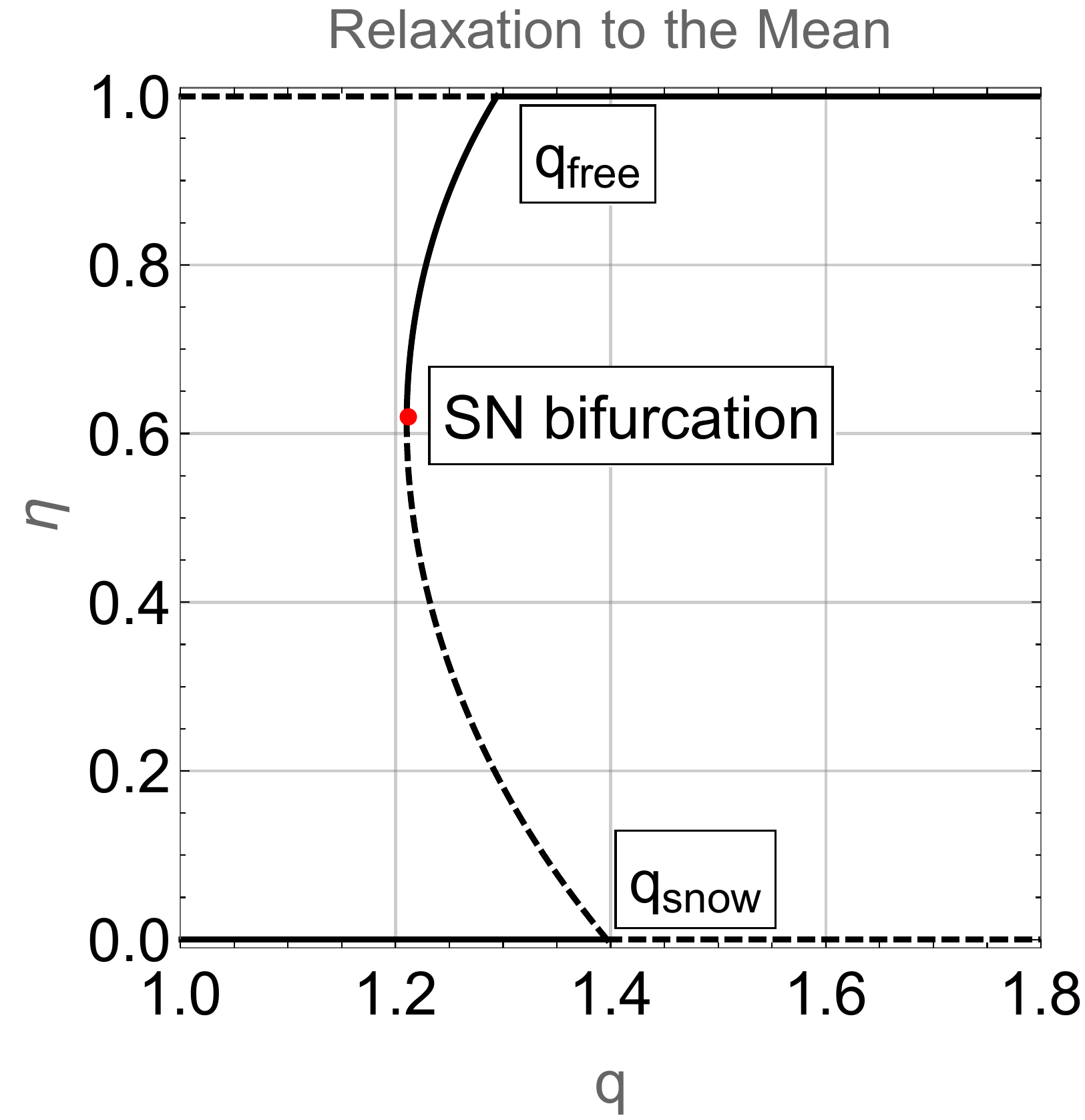} 
    \includegraphics[width=0.45\linewidth]{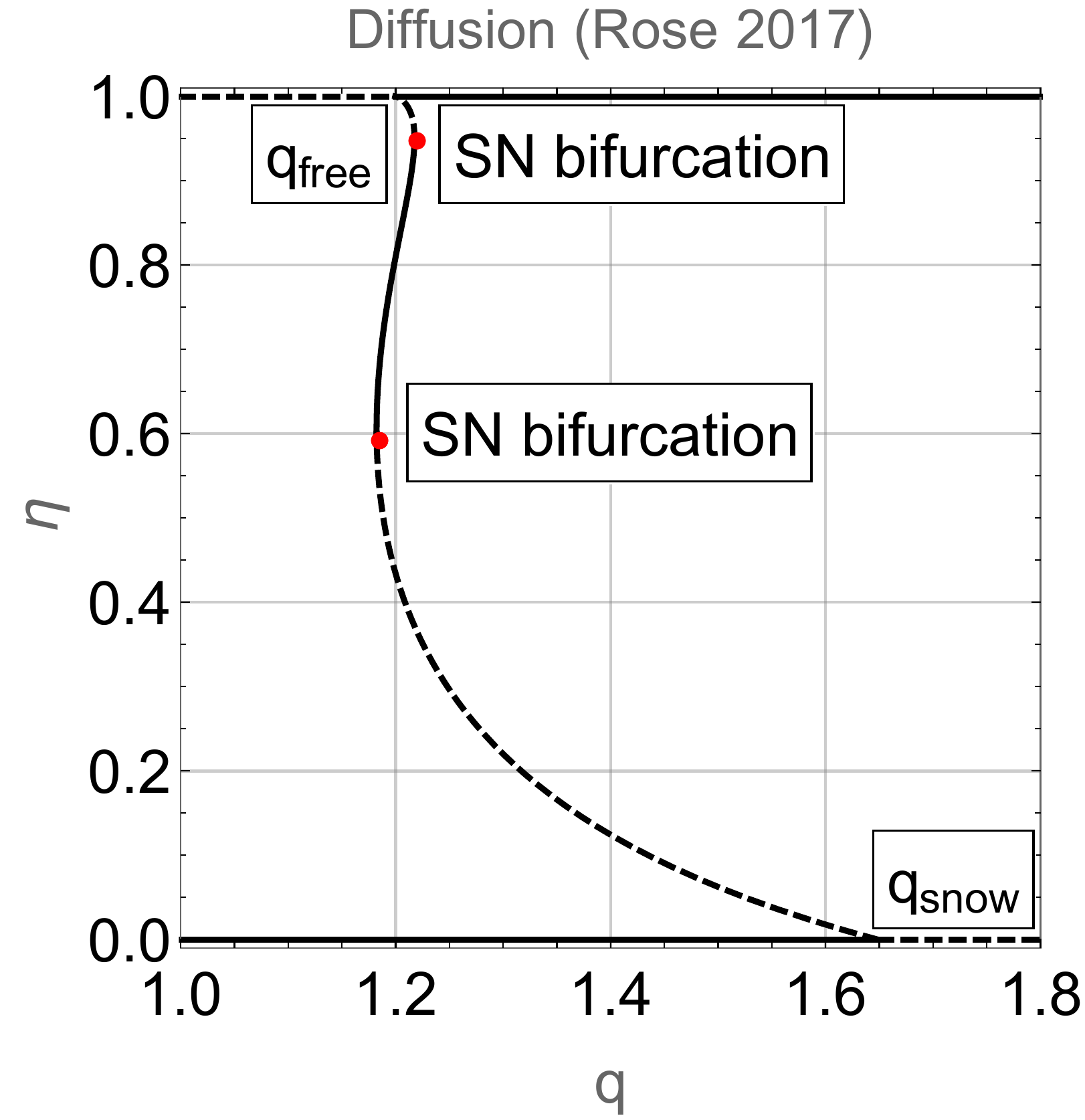}
  \caption{Plots showing the bifurcation diagrams demonstrating $q_{\text{free}}$, $q_{\text{snow}}$, and the saddle node point $\frac{\partial q_{\eta}}{\partial \eta}=0$ for Earth's parameter values from Table \ref{table:nondimensional_params} using the relaxation to the mean model used in this work (left) and the diffusion model used in \cite{Rose2017} (right). Solid curves indicate stable ice line locations. Dashed curves indicate unstable ice line locations. }
  \label{figure-2}
\end{figure}

For ice caps, a planet is ice free when $\eta=1$, and in a Snowball state when $\eta=0$.  Following the conventions from \cite{Rose2017}, for ice caps, we let $q_{\text{free}}$ denote the lowest value of $q$ for which $q_1$ is stable. Stability of the ice free state is inferred from where $q_\eta$ intersects the line $\eta=1$.  When $q>q_1$, the nondimensional temperature $T^{*}_{\text{eq}}$ at the pole is greater than 1. Similarly we let  $q_{\text{snow}}$ denote the location where $q_\eta$ intersects the line $\eta=0$.  We can think of this as the highest value of $q$ for which $q_0$ is stable. 
Note that for ice belts, $q_1$ is $q_{\text{snow}}$, and $q_0$ is $q_{\text{free}}$. 
See Figure \ref{figure-2}
for a bifurcation diagram demonstrating $q_{\text{free}}$ and $q_{\text{snow}}$. It should be noted that the definitions of $q_{\text{free}}$ and $q_{\text{snow}}$ used in this work take the average of the two sides of the discontinuity at the ice line. For ice caps, physically, the definition $q_{\text{free}}=q_1$ can be interpreted as vanishingly small ice caps, and $q_{\text{snow}}=q_0$ as vanishingly small equatorial strip of water. Alternative definitions can be used. We elaborate on the implications of this choice in Section \ref{sec:discussion}.

\subsubsection{Diffusion Equilibrium Temperature}

For the remainder, we will use the tilde symbol ($\tilde \ $) over a function or variable to denote that it is associated with the diffusion version of the model.

\cite{north1975analytical} analytically solves the diffusive energy balance equation \eqref{eq:nondim_model_diff} with $\partial T/\partial t=0$ for Earth with a degree two approximation to the insolation distribution.  He finds that solutions to the diffusive equation can be expressed in terms of hypergeometric functions.  This analytical solution is generalized to the exoplanet case with degree two approximation to the insolation distribution in \cite{Rose2017}.

For an arbitrary approximation to the insolation approximation, one may use the same methods described in \cite{north1975analytical} and \cite{Rose2017}.  For ease of notation, let

  \begin{align*}
   H_{2N}(x)= \sum_{n=0}^N\frac{a_{2n}p_{2n}(\zeta)}{1+2n(2n+1)\delta}p_{2n}(x).
\end{align*}   

As noted in \cite{north1975analytical}, $qH_{2N}(x)$ is the particular solution to the diffusion equation on the interval $[0,\eta)$ and  $q(1-\overline\alpha)H_{2N}(x)$ is the particular solution to the diffusion equation on the interval $(\eta,1]$. To find the general solution, one must find the solution to the homogeneous equation.  This derivation is given in \cite{north1975analytical}, so we do not give it here.  Instead we report the results of those computations which yield the equilibrium temperature at the ice line 
  \begin{align*}
\tilde T_{\text{eq}}^*(\eta)=\begin{cases}
q\left(H_{2N}(\eta) -\overline\alpha F_{2N}(\eta)\right),&\text{ice caps}\\
q\left((1-\overline\alpha)H_{2N}(\eta) +\overline\alpha F_{2N}(\eta)\right),&\text{ice belts}
\end{cases}
\end{align*}   
where
  \begin{align*}
    F_{2N}(\eta)&=\frac{P'_\nu(\eta) H_{2N}(\eta) -H'_{2N}(\eta)P_\nu(\eta)}{P'_\nu(\eta) f_\nu(\eta)-f'_\nu(\eta) P_\nu(\eta)}f_{\nu}(\eta),\\
    f_\nu(x)&= \ _2F_1(-\nu/2,(1+\nu)/2,1/2,x^2),\\
    P_\nu(x)&=\ _2F_1((1+\nu)/2,-\nu/2,1,1-x^2),\\
    \nu&=-\frac{1}{2}+\frac{(1-4/\delta)^{1/2}}{2},
\end{align*}   
and $\ _2F_1$ is the hypergeometric function.
Although $P_\nu(x)$ and $f_\nu(x)$ are complex-valued for $\delta<4$, taking the real parts of these functions yields linearly independent solutions as noted in \cite{north1975analytical}. Solving for $q$ gives
  \begin{align*}
\tilde q_\eta(\zeta,\overline\alpha,\delta)^{-1}=\begin{cases}
H_{2N}(\eta) -\overline\alpha F_{2N}(\eta),&\text{ice caps}\\
(1-\overline\alpha)H_{2N}(\eta) +\overline\alpha F_{2N}(\eta),&\text{ice belts}
\end{cases}
\end{align*}   
When $N=1$, the results from \cite{Rose2017} are recovered.

\section{Likelihood of Stable Partial Ice Cover}
\label{sec:likelihood}
\subsection{Defining the Region of Integration}
Planets with stable partial ice cover are potential candidates for a Snowball catastrophe bifurcation. We focus on quantifying the likelihood of stable partial ice cover depending on planetary obliquity. 
We follow \cite{Rose2017} and compute an estimate of the likelihood of stable partial ice cover based on the size of the region in parameter space that admits these types of solutions. 
Locations where $q_\eta$ or $\tilde q_\eta$ have a zero derivative demarcate regions of stable partial ice cover because the same location in the transformed plot with $\eta$ on the vertical axis is a saddle node bifurcation point (Figure \ref{figure-2}).

Taking the derivative of $q$ with respect to $\eta$ yields
  \begin{equation}
\begin{aligned}
   \frac{\partial q_{\eta}}{\partial \eta}  =&\frac{-2(1+\mu)}{[(2-\overline\alpha)\sigma_{2N}(\eta,\zeta)+2\mu  T_x(\eta)]^2} \times \\ &\quad\Big((2-\overline\alpha)\frac{\partial}{\partial \eta} \sigma_{2N}(\eta,\zeta) +2\mu (\pm \overline \alpha \sigma_{2N}(\eta,\zeta))\Big)
\label{EQ:dq}
\end{aligned}
\end{equation}   
for caps and belts,
where $\pm \overline \alpha \sigma_{2N}(\eta,\zeta)$ is the derivative of $T_x(\eta,\zeta)$ for caps and belts, respectively. Since $T_x(\eta)\geq0$, this definition is well defined. 
Parameter values that result in $\frac{\partial q_{\eta}}{\partial \eta}=0$ give the location of the saddle node. 
Setting Equation~\eqref{EQ:dq} to zero and solving for $\overline\alpha$ yields 
  \begin{equation}
\alpha_{\text{crit}}(\zeta,\mu,\eta)=
\begin{cases}
\frac{2\frac{\partial}{\partial \eta}\sigma_{2N}(\eta,\zeta)}{\frac{\partial}{\partial \eta}\sigma_{2N}(\eta,\zeta)-2 \mu \sigma_{2N}(\eta,\zeta)} \ \textnormal{for caps,} \\
\frac{2\frac{\partial}{\partial \eta}\sigma_{2N}(\eta,\zeta)}{\frac{\partial}{\partial \eta}\sigma_{2N}(\eta,\zeta)+2 \mu \sigma_{2N}(\eta,\zeta)} \ \textnormal{for belts}
\end{cases}
\label{eq:alpha_crit}
\end{equation}   
which is the critical value of the albedo contrast at the saddle node bifurcation latitude.  Stable partial ice cover is possible whenever $\overline\alpha<\alpha_{\text{crit}}(\zeta,\mu,\eta)$. At $\overline\alpha=\alpha_{\text{crit}}(\zeta,\mu,\eta)$, the saddle node bifurcation occurs.
Note that since $\mu$ is theoretically unbounded,  the function $\alpha_{\text{crit}}$ can become arbitrarily small.  In the next section, we use $\alpha_{\text{crit}}$ to find the relative likelihood of stable partial ice cover.

Following the method in \cite{Rose2017} but with the general degree $2N$ insolation approximation, we find that in the diffusion model.
  \begin{equation}
\tilde\alpha_{\text{crit}}(\zeta,\mu,\eta)=
\begin{cases}
\frac{H'_{2N}(\eta)}{F'_{2N}(\eta)} & \textnormal{for caps,} \\
\frac{H'_{2N}(\eta)}{H'_{2N}(\eta)-F'_{2N}(\eta)}  & \textnormal{for belts}.
\end{cases}
\label{eq:alpha_crit-diff}
\end{equation}   
Recall that $H_{2N}$ and $F_{2N}$ both also depend on $\zeta$ and $\delta$.

\subsection{Relative Likelihood of Stable Partial Ice Cover}

\begin{figure*}
  \centering

\includegraphics[width=0.95\linewidth]{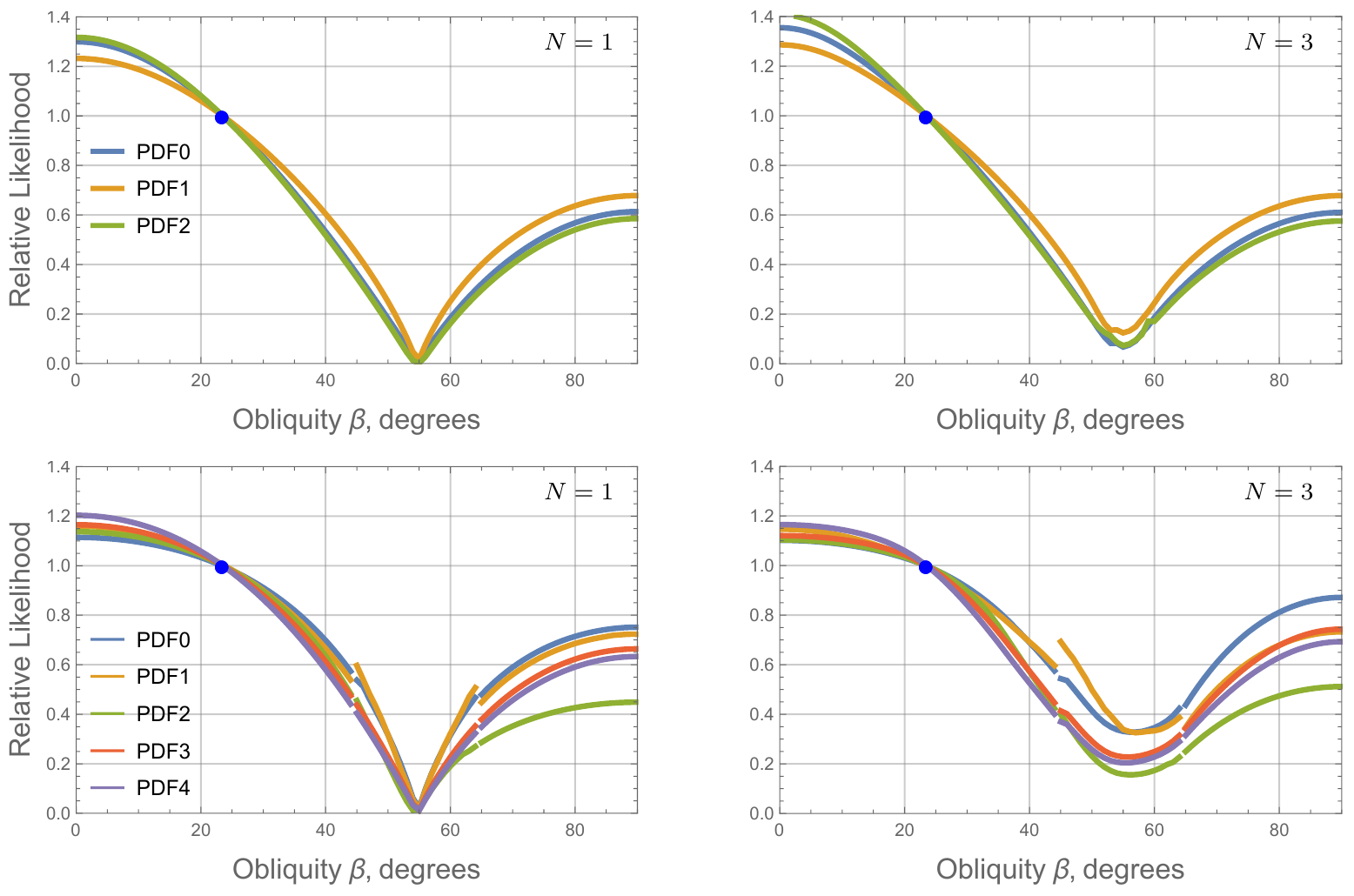} 
  \caption{Plots showing the relative likelihood of stable partial ice cover compared to partial stable ice cover on Earth. Plot show how the likelihood changes for different degree insolation approximations for First row: the diffusion model and Second row: the relaxation to the mean model. The top left plot is qualitatively similar to the likelihood plot in \cite{Rose2017}, but includes the likelihood contribution from inaccessible edges. Likelihoods are given for PDF0 (blue), PDF1 (orange), PDF2 (green) for both models and of PDF3 (red) and PDF4 (purple) for the relaxation to the mean model. The Earth obliquity is marked with a blue circle on each plot.}
  \label{figure-3}
\end{figure*}

In this section, we integrate over the parameter region in order to estimate the relative likelihood of stable partial ice cover using the same method as was described in \cite{Rose2017}.  We summarize it here for convenience.

We assume that there exists a true distribution for each parameter and that the parameters are independent of each other. However, since the true probability distributions are not known, we consider a handful of candidate probability distributions. We integrate the composite probability density function over the region of the domain where stable edges are present to obtain the likelihood of stable partial ice cover for a given value of obliquity $\zeta$. We normalize our results by the likelihood value for the obliquity of the Earth, $\zeta=\cos (23.5^{\circ})$. 
From the independence assumption, we can write the overall probability density function $h_{\text{planet}}$ as follows:

  \begin{equation*}
    h_{\text{planet}}=\begin{cases}
    h_q(q) h_{\mu}(\mu) h_{\overline\alpha}(\overline\alpha), & \text{relaxation to mean},\\
    h_q(q) h_{\delta}(\delta) h_{\overline\alpha}(\overline\alpha), & \text{diffusion}.
    \end{cases}
\end{equation*}   

We follow \cite{Rose2017} in our choice of candidate probability functions to test in order to facilitate the comparison with their results and include two additional probability functions to account for difference in $\delta$ and $\mu$. Since $q$, $\mu$, and $\delta$ are nonnegative and unbounded, log-normal  distributions are used to incorporate the possibility of a long tail. The general form of the probability density function of log-normal distribution is
  \begin{align*}
    h(x)=\frac{1}{\sigma \sqrt{2 \pi}(x-\theta)} \exp \left( \frac{-\left(\ln\left(\frac{x-\theta}{m}\right)\right)^2}{2\sigma^2}\right),
\end{align*}   

where $\sigma$ is referred to as the shape parameter, $m$ is the scale parameter, and $\theta$ is the location parameter. In our two new PDFs we use gamma distributions for $\mu$.  As with the log-normal distribution, the gamma distribution is a maximum entropy probability distribution and so minimizes the amount of prior information included in the distribution. The use of the gamma distribution allows us to create probability density functions that make Earth's value of $\mu=1.6$. While the gamma distribution has similar properties to those of the log-normal distribution, they are different enough to make them good candidates for sensitivity analysis \cite{wiens1999log, iaci2000gamma}.  The general form of the probability density function for the gamma distribution is
  \begin{align*}
    h(x)= \frac{b^ax^{a-1}e^{-bx}}{\Gamma(a)}
\end{align*}   
where $a$ is the shape parameter, $b$ is the rate parameter, and $\Gamma$ is the gamma function.  Since $\overline\alpha \in [0,1]$, uniform and beta distributions are used. The beta distribution has the probability density function
  \begin{align*}
h_{\overline\alpha}(\overline\alpha)=6\overline\alpha(1-\overline\alpha) 
\end{align*}   
and favors values of $\overline\alpha$ close to the value for Earth ($\overline\alpha=0.44$) compared to extremes of $\overline\alpha=0$ or 1. The probability density function parameters are chosen so that Earth's parameter values are not unlikely. 

For PDF0, $h_{\alpha}$ is uniform on $[0, 1]$; $h_{\mu}$ and $h_\delta$ are log-normal on $[0, \infty]$ with shape parameter 1.0, scale parameter 1.0, and location parameter 0; $h_q$ is log-normal on $[0, \infty]$ with shape parameter 0.5, scale parameter 1.0, and location parameter 0.

PDF1 is the same as PDF0 except $h_{\mu}$ and $h_\delta$ are log-normal on $[0, \infty]$ with shape parameter 2.0, scale parameter $e$, and location parameter 0. Compared to PDF0 and PDF2, PDF1 makes very small values and large values of $\mu$ or $\delta$ more likely.

PDF2 is the same as PDF0 except $h_{\alpha}$ is parabolic beta-distribution on $[0, 1]$, with mode at $0.5$. Compared to PDF0 and PDF1, PDF2 favors intermediate values of albedo contrast compared to extreme values. 

PDF3 is the same as PDF0 except $h_{\mu}$ and $h_\delta$ have a gamma distribution on $[0, \infty]$ with shape parameter 4.0, rate parameter 2.5.  Compared to PDF0, PDF3 makes the $\mu$ value for Earth (1.6) more likely.

PDF4 is the same as PDF1 except $h_{\mu}$ and $h_\delta$ have a gamma distribution on $[0, \infty]$ with shape parameter 4.0, rate parameter 1.85.  Compared to PDF3, PDF4 makes larger values of $\mu$ more likely.

The (non-normalized) likelihood is given by
\begin{equation}
    P_{\text{ice}}(\beta)=\frac{\int_0^1 \int_0^{\infty}\int_0^{\alpha_{\text{crit}}} h_{\text{planet}} (q_{\eta},\mu,\alpha) d\alpha \ d\mu \ d\eta}{\int_0^{\infty} \int_0^{\infty}\int_0^{1} h_{\text{planet}} (q,\mu,\alpha) d\alpha \ d\mu\ dq }. 
    \label{eq:likelihood}
\end{equation}
Note that the denominator in equation \eqref{eq:likelihood} is equal to 1 since $h_{\text{planet}}$ is a probability density function. 
%The integration region is illustrated in Figure \ref{fig:integr_region}. 
For higher values of $\mu$ and $\delta$, the corresponding values of $\alpha_{\text{crit}}$ and $\alpha_{\text{crit}}$ become vanishingly small for all latitudes $\eta$. Therefore, even though the region of integration is unbounded, numerical integration converges. The results of the integration are summarized in Figure \ref{figure-3}.

\begin{figure}
  \centering
\includegraphics[width=0.99\linewidth]{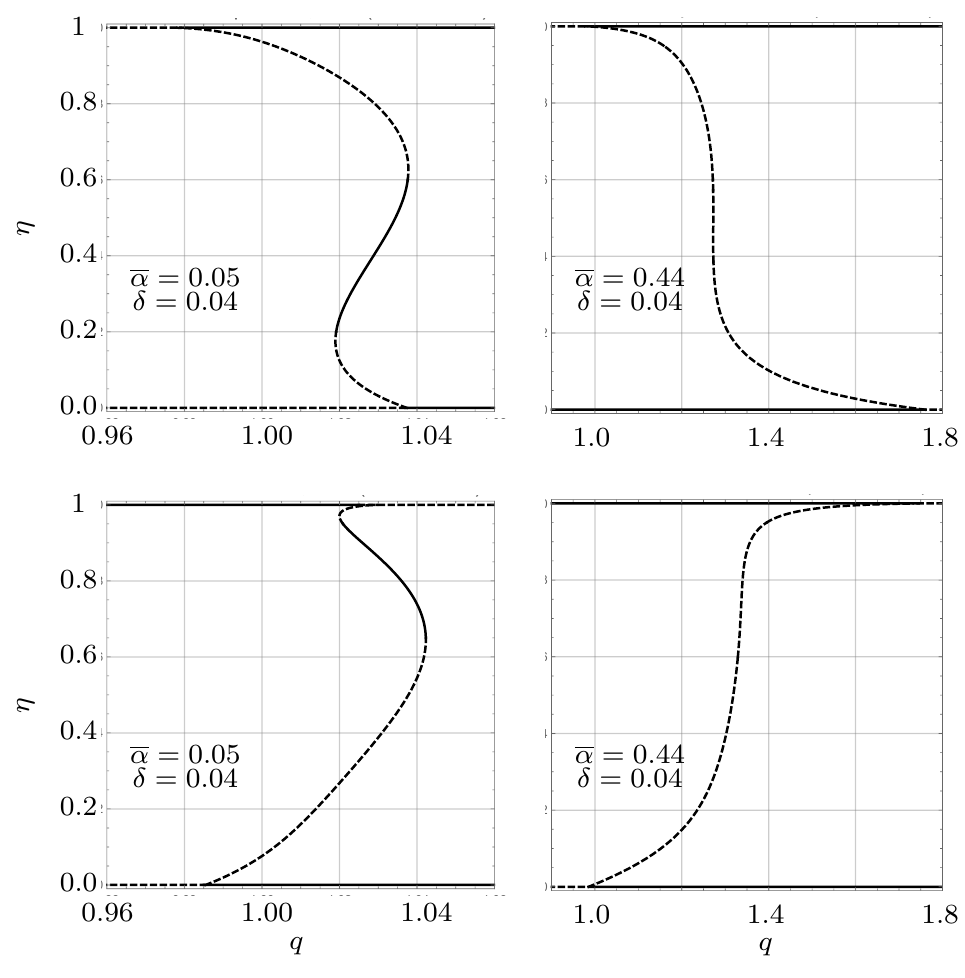} 
  \caption{Plots showing stability of the equilibrium ice line locations for $\beta=54.47^\circ$ for the diffusion model for ice caps (top row) and ice belt (bottom row).  The model takes $N=3$, the minimum approximation needed to capture the qualitative characteristics of the insolation distribution at this obliquity. Solid curves indicate stable ice line locations. Dashed curves indicate unstable ice line locations. Note that the horizontal scales are different between the left and right columns.}

  \label{figure-4}
\end{figure}

The qualitative results are similar across all models. Planets at low obliquity have a higher likelihood of stable partial ice cover than planets at high obliquity, and planets with moderate obliquity have the lowest likelihood of stable partial ice cover.   The apparent discontinuities in the relative likelihood plots for the relaxation to the mean model in Figure \ref{figure-3} are due to cut-offs for the obliquities where we use a model with caps or belts. As shown by \cite{dobrovolskis2020}, the minima of the insolation distribution occur between the poles and the equator when the obliquity is between approximately $45^\circ$ and $65.355^\circ$.
This behavior can be thought of as a transitional regime between ice caps (where the insolation minima occur at the poles) and ice belts (where the insolation minimum occurs at the equator). While we do not specifically consider planets with varying obliquity in this work, we use both the caps and belts models for this range of obliquities.

In agreement with the degree two diffusion model in \cite{Rose2017}, the likelihood of stable partial ice cover goes to zero for the degree 2 relaxation to the mean model. It is easy to see the reason for this by noting that the derivative of the insolation function appears in the numerator of $\alpha_{\text{crit}}$ and recalling that the insolation function $\sigma_2$ is constant for $\zeta=\sqrt{3}/3$ which is when the obliquity is $54.74^\circ$.  
For the models with higher degree insolation approximations, the likelihood is never identically zero but nevertheless attains its minimum at mid-obliquities for all tested probability distribution functions. 

For example, in Figure \ref{figure-4}  we plot bifurcation diagrams for the diffusion model at $\beta=54.47^\circ$. For $N=3$ shown in the figure, small values of $\overline\alpha$ and $\delta$ admit stable partial ice edges for both ice caps and ice belts, although it is only for the smallest values of these parameters that the edges are accessible in a hysteresis loop. Once $\alpha\approx0.3$, there are no longer any stable partial ice edges in the diffusion model. In contrast, no value of albedo contrast nor efficiency of heat transport will cause the diagram for $N=1$ to have a saddle node bifurcation.

\section{Partial Ice Cover and the Snowball State}
\label{sec:partial-ice-snowball}

\begin{figure*}
  \centering
\includegraphics[width=0.99\linewidth]{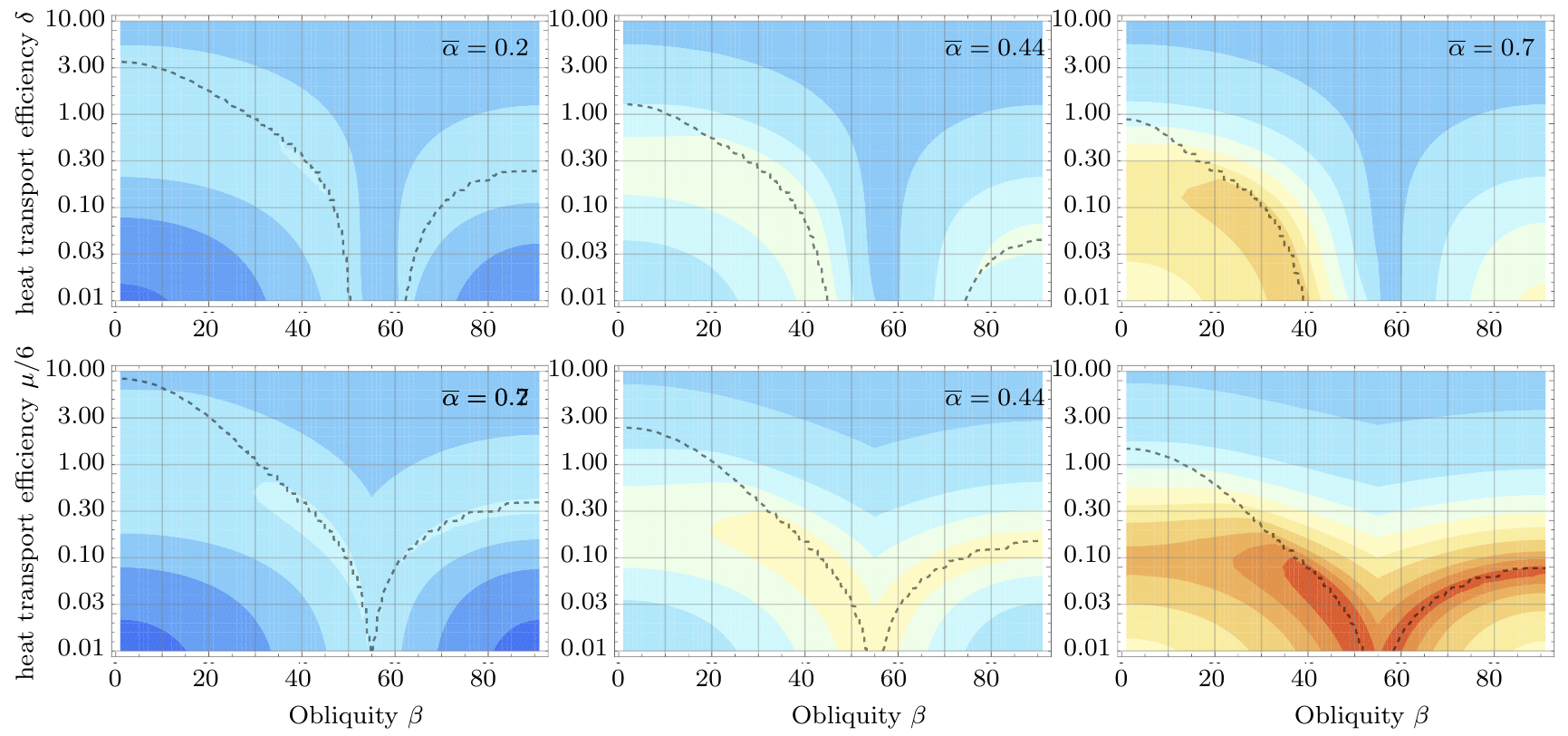} \\
  \includegraphics[width=0.6\linewidth]{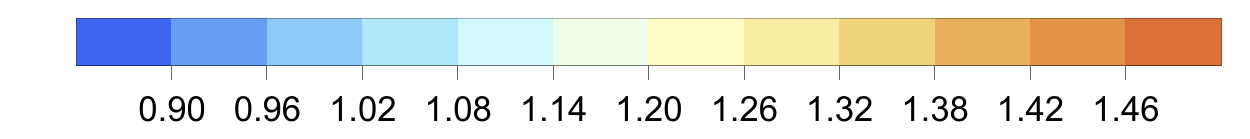}
  \caption{Contour plots showing the minimum value of the nondimensional incoming radiation $q$ for which the Snowball state is not globally stable. Regions above the dashed curves are where the Snowball catastrophe occurs directly from the ice free state. \textbf{First row:} The diffusion model with $N=3$. \textbf{Second row:} The relaxation to the mean model with $N=3$. }
  \label{figure-5}
\end{figure*}

In addition to assessing the likelihood of partial ice cover, we quantify the severity of the Snowball state. Typically, one is interested in determining the inner and outer edges of the habitable zone based on the system parameters; however, as \cite{Rose2017} notes in Section 5.3 (and references therein), the models are too simplistic to give good estimates for this range. It is possible to quantify whether a bifurcation from no or partial ice cover to the Snowball state occurs for the system parameters.

The most severe Snowball bifurcation occurs when $q_{\text{free}}<q_{\text{snow}}$ (similarly $\tilde q_{\text{free}}<\tilde q_{\text{snow}}$), and for all $0<\eta <1 $, $q_{\text{free}}\leq q_{\eta}\leq  q_{\text{snow}}$. Although there may be stable ice line equilibria between 0 and 1, they would inaccessible by varying $q$ through a hysteresis loop in this situation.
A less severe bifurcation occurs when the ice line continuously transitions from the ice free state to small stable ice caps (or ice belt) and then the ice line drops to the Snowball state. Behavior of solutions is the typical passage through a saddle node bifurcation \cite{strogatz2018nonlinear}.  The least severe scenario is where no saddle node bifurcation occurs and stable ice line equilibrium depends continuously on the parameter $q$. This occurs when $q_{\text{snow}}<q_{\text{free}}$, and for all  $0<\eta <1 $, $q_{\text{snow}}\leq q_{\eta} \leq q_{\text{free}}$.

In Figure \ref{figure-5}, we plot contours for the smallest value of the nondimensional incoming radiation $q$ for which the Snowball state is not globally attracting (i.e. the only stable solution).  In \cite{Rose2017}, this value was referred to as $q_{\text{hab}}$ and computed by
  \[q_{\text{hab}}=\min\{q_{\text{free}},q_{\text{stab}}\}\]   
where $q_{\text{stab}}$ is the minimum $q$ value for which an ice line equilibrium exists between 0 and 1.  Frequently this corresponds to the saddle node bifurcation which causes the Snowball catastrophe (if present, e.g. in Figure \ref{figure-4} only the lower left plot exhibits this bifurcation). If the model has stable partial ice edges between 0 and 1 but no saddle node bifurcation causing the Snowball catastrophe, $q_{\text{stab}}=q_{\text{snow}}$. %(e.g. in Figure \ref{figure-4} the bottom two plots have this feature).
When $q_{\text{free}}>q_{\text{stab}}$, the hysteresis loop generated by varying $q$ causes a drop from the ice free state to a partially ice covered state instead of directly to the Snowball state.

In Figure \ref{figure-5}, the regions above the dashed black curves are where the Snowball catastrophe occurs directly from the ice free state.
In an extension of the work in \cite{Rose2017}, here we consider the models with the degree six approximation to the insolation distribution instead of the degree two approximation for insolation in the diffusion model.  For ease of comparison with the results in \cite{Rose2017}, we restrict the ice cap models to obliquities between $0^\circ$ and $55^\circ$ and the ice belt models to obliquities between $55^\circ$ and $90^\circ$.

It is straightforward to compare the top row of Figure \ref{figure-5} to Figure 8 in \cite{Rose2017}. Analytically, the only difference is the degree of the insolation distribution used.  On the whole, the figures are very similar, having nearly the same contours.  They differ only slightly in the maximum and minimum values: the sixth degree approximation decreases the extrema achieved for $q_{\text{hab}}$ for the model with ice caps and increases the extrema for the model with an ice belt relative to the second degree approximation.  This is due to the increased accuracy of the insolation approximation at the poles (minima for ice caps, maxima for an ice belt). The largest difference is that for the smallest obliquities in the degree six diffusion model, the Snowball catastrophe from the ice free state occurs for larger values of the heat transport parameter. This means the Snowball catastrophe is less severe for low obliquity planets than in the degree two diffusion model.  The transition from ice free directly to the Snowball state is minimally affected for obliquities above approximately $30^\circ$. 

In the bottom row of Figure \ref{figure-5}  we plot the contours for the degree six relaxation to the mean model.  The relaxation to the mean model has stark contrasts with the diffusion model in the row above it.  For low to moderate albedo contrast the minima and maxima of both models are relatively similar. However for large albedo contrast, the maximum values achieved for the relaxation to the mean model are significantly greater. Analytically this is caused by the dependence of $q_{\text{free}}$ on the albedo contrast.  In the diffusion model  $q_{\text{free}}$ is independent of $\overline\alpha$, but in the relaxation to the mean model it is not. As $\overline \alpha$ increases so does $q_{\text{free}}$ for the relaxation to the mean model (this can be seen in equation \eqref{eq:q_eta}). %(see for example Figure \ref{figure-4}).

In both models, increasing the albedo contrast increases the severity of the Snowball catastrophe for all obliquities (the dashed black curves decrease), but the increase in severity in the relaxation to the mean model is minimal. Although for all obliquities in either model lower heat transport efficiency always guards against the most severe transition (from ice free directly to Snowball). In the diffusion model for the range of heat transport plotted this most severe transition is always present for the mid-obliquities and increasingly present for higher obliquities as the albedo contrast increases. To avoid the ice free directly to Snowball transition requires decreasing the heat transport efficiency at least another order of magnitude than what is plotted (i.e. to at least 0.001 for $\beta=90^\circ$ and $\overline \alpha=0.7$). Conversely, in the relaxation to the mean model there is only a small range of obliquities where this most severe transition is present for the range of heat transport considered.

\section{Discussion}
\label{sec:discussion}

In this paper we have analyzed a one dimensional energy balance model with heat transport modeled by relaxation to the global mean temperature and diffusion. The  relaxation to the mean and diffusion versions of the Budyko-Sellers
model provide two similar ways to model the multitude of processes involved in energy transfer between latitudes. Both methods ensure that energy is transported from latitudes that are ``hot'' to ones that are ``cold.'' The diffusive heat transport is a local process that necessitates special treatment at the poles, while relaxation to the mean global temperature is a global process that does not require special boundary conditions \cite{widiasih2013dynamics}.  This work may be interpreted for any rapidly rotating rocky planet with some physical mechanism of heat transport.  This work applies to planets where the temperature affects the albedo, in particular, we assume that higher temperatures decrease the albedo as they do for ice/water on Earth.

We have also considered the effects of different approximations to the annual average insolation distribution on the results of the models. The second degree approximation of the insolation distribution used in \cite{Rose2017} does not capture the qualitative distribution of mid-obliquity planets.  Planets with obliquities between approximately $45^\circ$ and $65^\circ$ have a characteristic `W' shape that requires a degree six (or higher) polynomial approximation \cite{Nadeau2017}.  

A main result from \cite{Rose2017} is that the likelihood of stable partial ice cover goes to zero at $55^{\circ}$ obliquity. This result is due to the insolation approximation used in their study, which is constant for $55^{\circ}$ obliquity.  In the definition of $\tilde \alpha_{\text{crit}}$ (equation \eqref{eq:alpha_crit-diff}), if $H_{2N}$ were constant then $\tilde \alpha_{\text{crit}}=0$ for all values of the arguments.  When $N=1$ as in \cite{Rose2017}, the function $H_{2N}$ is constant exactly when the insolation is constant, i.e. when the obliquity is $54.47^\circ$. This means that the integral in  the numerator of the likelihood calculation (equation \eqref{eq:likelihood}) is zero.  We see a similar problem with the relaxation to the mean model when $N=1$.  Here it is clear to see that when the insolation is constant then $\alpha_{\text{crit}}=0$ (see Equation~\eqref{eq:alpha_crit}). Taking a higher degree approximation in either model, as we do here, avoids these problems.

In Figure \ref{figure-3} we saw that low obliquity planets are more likely to have stable partial ice cover than those with high obliquity, and planets at middle obliquities are least likely to have stable partial ice cover, which is qualitatively similar to the likelihood computations in \cite{Rose2017}. As noted above, we find that stable partial ice cover is possible at all obliquities and that, in particular, the relative likelihood of finding a planet with partial stable ice cover ranges between 7\% and 32\%.

Comparing the relaxation to the mean model to the diffusion model, we note that the latter predicts lower likelihood of stable partial ice cover at lower obliquities. This effect is due to the fact that the diffusion model can exhibit a second saddle node bifurcation at high values of $\eta$, close to the poles (called the small ice cap instability) for Earth's obliquity. As the obliquity decreases to zero   the  small  ice  cap  instability shrinks in size and can disappear completely for small to moderate values of the albedo contrast $\overline\alpha$, increasing the range of $\eta$ where stable ice edges are possible. 
In contrast, the relaxation to the mean model does not exhibit the small ice cap instability at Earth's obliquity for degree two or degree six insolation approximation. This means that the likelihood remains relatively flat as the obliquity decreases.

The relaxation to the mean model also exhibits pronounced differences between PDF0, PDF1 and PDF2 at high obliquities. For high values of the obliquity $\beta$, the likelihood of stable partial ice cover is lower for PDF2. The difference between PDF2 and other tested probability density functions is due to the differences in $\alpha_{\text{crit}}$ between the relaxation to the mean model and the diffusion model. Since PDF2 changes the distribution of $\overline \alpha$ from a uniform distribution to a parabolic beta distribution, the shape of $\alpha_{\text{crit}}$ results in a more pronounced difference for the relaxation to the mean model than for the diffusion model. The resulting gap between the likelihood curves conveys decreased certainty about the likelihood of stable partial ice cover on high obliquity planets.  Note also that for PDF3 and PDF4, the relaxation to the mean model exhibits behavior similar to that for PDF1.

We quantify the effects of albedo contrast and efficiency of heat transport on the presence of hysteresis loops in radiative forcing.  We find that the severity of the Snowball bifurcation increases as the albedo contrast $\overline\alpha$ and the efficiency of heat transport ($\delta$ or $\mu$) increase.

The above behavior can be explained by the effect of albedo contrast and efficiency of heat transport on the planetary climate mechanism. When the albedo contrast $\overline\alpha$ is low, ice is not much more reflective than the non-frozen regions (either ground or water), resulting in suppressed ice--albedo feedback. When $\delta$ or $\mu$ are low, the near-absence of heat transport across latitudes limits the interaction between the ice regions and the water regions of the planet, thus reducing the likelihood of the Snowball catastrophe. When  $\overline\alpha$ is high, the reflectivity of ice is much higher than that of the non-frozen regions, expediting the ice---albedo processes. When $\mu$ is high, the heat transport across latitudes makes it difficult to maintain a difference in temperatures between ice regions and water regions, leading to an ice free or a Snowball planet. 

In the range of obliquities where both ice caps and ice belts may be stable it is not possible to transition continuously between the ice free and Snowball states as $q$ is varied, i.e. there will always be a hysteresis loop when varying $q$.  The hysteresis will either contain a saddle node bifurcation or the most severe Snowball bifurcation from ice free to completely ice covered. A future study might explore whether the lack of a region without hysteresis in the parameter space is a contributing factor for the decrease in the likelihood of stable partial ice cover for these obliquities in Figure \ref{figure-3}.

It should be noted that dealing with the discontinuity at the ice line affects the definitions of $q_{\text{free}}$ and $q_{\text{snow}}$.  For ice caps, physically, the definition $q_{\text{free}}=q_1$ can be interpreted as vanishingly small ice caps, and $q_{\text{snow}}=q_0$ as vanishingly small equatorial strip of water. This choice is consistent with the literature on the relaxation to the mean model \cite{widiasih2013dynamics,mcgehee2014}. This definition ensures that the bifurcation diagram for the relaxation to the mean model is continuous at $\eta=0$ and $\eta=1$. However, one could consider an alternative definition of $q_{\text{free}}$ and $q_{\text{snow}}$: instead of taking the average of the two sides of the discontinuity at the ice line, one could use only the warm branch (Equation \eqref{eq:warm-branch}) of the temperature equilibrium to define $q_{\text{free}}$ and only the cold branch (Equation \eqref{eq:cold-branch}) to define $q_{\text{snow}}$. This choice yields $q_{\text{free}}=\frac{(1+\mu)}{\sigma_{2N}(\eta,\zeta)+\mu}$ and $q_{\text{snow}}=\frac{(1+\mu)}{(1-\overline{\alpha})(\sigma_{2N}(\eta,\zeta)+\mu)}$. Such a definition might be more physically intuitive, representing the relevant thresholds for the first appearance of freezing temperatures at the poles for a warm planet undergoing a gradual cooling process, and the first appearance of thawing temperatures for a planet undergoing a gradual thawing process. Removing the dependence of $q_{\text{free}}$ on $\overline{\alpha}$ might lead to a higher estimate for the likelihood of ice-free solutions. However, analyzing the implications of these alternatives would require extensive mathematical analysis of bistability of ice free and partial ice states that we defer to future work.

The robustness of the Snowball catastrophe and the parameter regimes where an energy balance model might be applicable has been debated. In the GCM simulations conducted by \cite{ferreira2014climate}, the Snowball catastrophe occurs only for particular ocean regimes.  \cite{wagner2015climate} show that meridional heat transfer may increase ice cover stability. \cite{rose2009ocean} have extended the energy balance models to include ocean heat transport and meridional structure and have found that Snowball catastrophe is possible in those models. 

If this work were to be applied to an observed planet, the obliquity $\zeta$ and the albedo contrast $\overline \alpha$ could perhaps be measured, and albedo signatures could be compared to the predictions of our model. The parameter $q$ will be more difficult to estimate. While the mean annual insolation $Q$ can be derived from information about the star and orbit of the planet, the dependence of $q$ on the atmospheric parameters $A$ and $B$ make determining $q$ challenging. The parameters $\delta$ and $\mu$, efficiency of heat transport, would also be difficult to measure directly because of our limited knowledge of rates of heat transport on different planets.

\section{Conclusion}
\label{sec:conclusion}

In this paper we have analyzed a one dimensional energy balance model with two different methods of modeling the heat transport.  The models have explicit dependence on the planet's obliquity through the annual average insolation distribution which we approximate with degree two and degree six polynomials. We pay particular attention to the planet's obliquity, radiative forcing,  albedo contrast, and efficiency of heat transport and find:
\begin{enumerate}
    \item With an improved approximation to the insolation distribution function, planets at all values of obliquity exhibit nonzero likelihood of stable partial ice cover regardless of mode of heat transport. Minimum likelihood ranges from 7\% to 32\% relative to Earth's likelihood of stable partial ice cover, depending on the model and probability density functions for the parameters. Maximum likelihood ranges from 110\% to 140\% relative to Earth.
    \item Several results from the earlier study by \cite{Rose2017} are seen here in both the diffusion and relaxation to the mean models:
        \begin{enumerate}
            \item Models with high obliquity  are less likely to have stable partial ice cover than ones with low obliquity but are more likely than models with moderate obliquity.
            \item Low albedo contrast and low efficiency of heat transport favor stable partial ice cover.
            \item High albedo contrast and high efficiency of heat transport favor severe Snowball catastrophe in a hysteresis loop caused by changes in radiative forcing.
        \end{enumerate}
    \item In the relaxation to the mean model, a larger range of heat transport efficiency supports partial stable ice cover at high obliquities even for high albedo contrast. Whereas in the diffusion model stable partial ice cover a high obliquities and high albedo contrast is possible only for very low heat transport efficiency.
\end{enumerate}

%include exoplanets (a general planet) end with a sentence or two.

Both models discussed in this work are highly simplified, and so direct implications for real physical systems are tenuous when taken individually. We may interpret the results with higher confidence in regions of parameter space where their solutions give similar results. Places where they differ indicate need for further investigations with more complex models.

For example, due to the disagreement of the models shown in Figure \ref{figure-5}, a future study may explore the transition to the Snowball state for high obliquity planets in a general circulation model. Other extensions of this work would be to look at climatic histories of planets by introducing obliquity or eccentricity variations in time into the energy balance model used in this study. Variations in a planet's orbital parameters change the behavior of planetary ice cover over geological time and which are easiest to model over long time periods with simple energy balance models like the ones presented here.

\acknowledgements{
The authors thank Nikole Lewis, Tiffany Kataria, Toby Ault, and Steven Strogatz for their comments in preparing the manuscript. The authors also thank Brian Rose and an anonymous reviewer for their comments during the review process which greatly improved the manuscript. AN was supported by Mathematical Sciences Postdoctoral Research Fellowship  (Award Number DMS-1902887) during this project.}

\bibliography{main_v3}

\begin{thebibliography}{35}%
\makeatletter
\providecommand \@ifxundefined [1]{%
 \@ifx{#1\undefined}
}%
\providecommand \@ifnum [1]{%
 \ifnum #1\expandafter \@firstoftwo
 \else \expandafter \@secondoftwo
 \fi
}%
\providecommand \@ifx [1]{%
 \ifx #1\expandafter \@firstoftwo
 \else \expandafter \@secondoftwo
 \fi
}%
\providecommand \natexlab [1]{#1}%
\providecommand \enquote  [1]{``#1''}%
\providecommand \bibnamefont  [1]{#1}%
\providecommand \bibfnamefont [1]{#1}%
\providecommand \citenamefont [1]{#1}%
\providecommand \href@noop [0]{\@secondoftwo}%
\providecommand \href [0]{\begingroup \@sanitize@url \@href}%
\providecommand \@href[1]{\@@startlink{#1}\@@href}%
\providecommand \@@href[1]{\endgroup#1\@@endlink}%
\providecommand \@sanitize@url [0]{\catcode `\\12\catcode `\$12\catcode
  `\&12\catcode `\#12\catcode `\^12\catcode `\_12\catcode `\%12\relax}%
\providecommand \@@startlink[1]{}%
\providecommand \@@endlink[0]{}%
\providecommand \url  [0]{\begingroup\@sanitize@url \@url }%
\providecommand \@url [1]{\endgroup\@href {#1}{\urlprefix }}%
\providecommand \urlprefix  [0]{URL }%
\providecommand \Eprint [0]{\href }%
\providecommand \doibase [0]{http://dx.doi.org/}%
\providecommand \selectlanguage [0]{\@gobble}%
\providecommand \bibinfo  [0]{\@secondoftwo}%
\providecommand \bibfield  [0]{\@secondoftwo}%
\providecommand \translation [1]{[#1]}%
\providecommand \BibitemOpen [0]{}%
\providecommand \bibitemStop [0]{}%
\providecommand \bibitemNoStop [0]{.\EOS\space}%
\providecommand \EOS [0]{\spacefactor3000\relax}%
\providecommand \BibitemShut  [1]{\csname bibitem#1\endcsname}%
\let\auto@bib@innerbib\@empty
%</preamble>
\bibitem [{\citenamefont {Armstrong}\ \emph {et~al.}(2014)\citenamefont
  {Armstrong}, \citenamefont {Barnes}, \citenamefont {Domagal-Goldman},
  \citenamefont {Breiner}, \citenamefont {Quinn},\ and\ \citenamefont
  {Meadows}}]{armstrong2014effects}%
  \BibitemOpen
  \bibfield  {author} {\bibinfo {author} {\bibfnamefont {J.}~\bibnamefont
  {Armstrong}}, \bibinfo {author} {\bibfnamefont {R.}~\bibnamefont {Barnes}},
  \bibinfo {author} {\bibfnamefont {S.}~\bibnamefont {Domagal-Goldman}},
  \bibinfo {author} {\bibfnamefont {J.}~\bibnamefont {Breiner}}, \bibinfo
  {author} {\bibfnamefont {T.}~\bibnamefont {Quinn}}, \ and\ \bibinfo {author}
  {\bibfnamefont {V.}~\bibnamefont {Meadows}},\ }\href@noop {} {\bibfield
  {journal} {\bibinfo  {journal} {Astrobiology}\ }\textbf {\bibinfo {volume}
  {14}},\ \bibinfo {pages} {277} (\bibinfo {year} {2014})}\BibitemShut
  {NoStop}%
\bibitem [{\citenamefont {Rose}\ \emph {et~al.}(2017)\citenamefont {Rose},
  \citenamefont {Cronin},\ and\ \citenamefont {Bitz}}]{Rose2017}%
  \BibitemOpen
  \bibfield  {author} {\bibinfo {author} {\bibfnamefont {B.~E.}\ \bibnamefont
  {Rose}}, \bibinfo {author} {\bibfnamefont {T.~W.}\ \bibnamefont {Cronin}}, \
  and\ \bibinfo {author} {\bibfnamefont {C.~M.}\ \bibnamefont {Bitz}},\
  }\href@noop {} {\bibfield  {journal} {\bibinfo  {journal} {The Astrophysical
  Journal}\ }\textbf {\bibinfo {volume} {846}},\ \bibinfo {pages} {28}
  (\bibinfo {year} {2017})}\BibitemShut {NoStop}%
\bibitem [{\citenamefont {Kane}\ \emph {et~al.}(2020)\citenamefont {Kane},
  \citenamefont {Li}, \citenamefont {Wolf}, \citenamefont {Ostberg},\ and\
  \citenamefont {Hill}}]{kane2020eccentricity}%
  \BibitemOpen
  \bibfield  {author} {\bibinfo {author} {\bibfnamefont {S.~R.}\ \bibnamefont
  {Kane}}, \bibinfo {author} {\bibfnamefont {Z.}~\bibnamefont {Li}}, \bibinfo
  {author} {\bibfnamefont {E.~T.}\ \bibnamefont {Wolf}}, \bibinfo {author}
  {\bibfnamefont {C.}~\bibnamefont {Ostberg}}, \ and\ \bibinfo {author}
  {\bibfnamefont {M.~L.}\ \bibnamefont {Hill}},\ }\href@noop {} {\bibfield
  {journal} {\bibinfo  {journal} {The Astronomical Journal}\ }\textbf {\bibinfo
  {volume} {161}},\ \bibinfo {pages} {31} (\bibinfo {year} {2020})}\BibitemShut
  {NoStop}%
\bibitem [{\citenamefont {Checlair}\ \emph {et~al.}(2017)\citenamefont
  {Checlair}, \citenamefont {Menou},\ and\ \citenamefont
  {Abbot}}]{checlair2017no}%
  \BibitemOpen
  \bibfield  {author} {\bibinfo {author} {\bibfnamefont {J.}~\bibnamefont
  {Checlair}}, \bibinfo {author} {\bibfnamefont {K.}~\bibnamefont {Menou}}, \
  and\ \bibinfo {author} {\bibfnamefont {D.~S.}\ \bibnamefont {Abbot}},\
  }\href@noop {} {\bibfield  {journal} {\bibinfo  {journal} {The Astrophysical
  Journal}\ }\textbf {\bibinfo {volume} {845}},\ \bibinfo {pages} {132}
  (\bibinfo {year} {2017})}\BibitemShut {NoStop}%
\bibitem [{\citenamefont {Checlair}\ \emph
  {et~al.}(2019{\natexlab{a}})\citenamefont {Checlair}, \citenamefont {Olson},
  \citenamefont {Jansen},\ and\ \citenamefont {Abbot}}]{checlair2019ocean}%
  \BibitemOpen
  \bibfield  {author} {\bibinfo {author} {\bibfnamefont {J.~H.}\ \bibnamefont
  {Checlair}}, \bibinfo {author} {\bibfnamefont {S.~L.}\ \bibnamefont {Olson}},
  \bibinfo {author} {\bibfnamefont {M.~F.}\ \bibnamefont {Jansen}}, \ and\
  \bibinfo {author} {\bibfnamefont {D.~S.}\ \bibnamefont {Abbot}},\ }\href@noop
  {} {\bibfield  {journal} {\bibinfo  {journal} {The Astrophysical Journal
  Letters}\ }\textbf {\bibinfo {volume} {884}},\ \bibinfo {pages} {L46}
  (\bibinfo {year} {2019}{\natexlab{a}})}\BibitemShut {NoStop}%
\bibitem [{\citenamefont {Checlair}\ \emph
  {et~al.}(2019{\natexlab{b}})\citenamefont {Checlair}, \citenamefont
  {Salazar}, \citenamefont {Paradise}, \citenamefont {Menou},\ and\
  \citenamefont {Abbot}}]{checlair2019HZ}%
  \BibitemOpen
  \bibfield  {author} {\bibinfo {author} {\bibfnamefont {J.~H.}\ \bibnamefont
  {Checlair}}, \bibinfo {author} {\bibfnamefont {A.~M.}\ \bibnamefont
  {Salazar}}, \bibinfo {author} {\bibfnamefont {A.}~\bibnamefont {Paradise}},
  \bibinfo {author} {\bibfnamefont {K.}~\bibnamefont {Menou}}, \ and\ \bibinfo
  {author} {\bibfnamefont {D.~S.}\ \bibnamefont {Abbot}},\ }\href@noop {}
  {\bibfield  {journal} {\bibinfo  {journal} {The Astrophysical Journal
  Letters}\ }\textbf {\bibinfo {volume} {887}},\ \bibinfo {pages} {L3}
  (\bibinfo {year} {2019}{\natexlab{b}})}\BibitemShut {NoStop}%
\bibitem [{\citenamefont {Yue}\ and\ \citenamefont
  {Yang}(2020)}]{yue2020effect}%
  \BibitemOpen
  \bibfield  {author} {\bibinfo {author} {\bibfnamefont {W.}~\bibnamefont
  {Yue}}\ and\ \bibinfo {author} {\bibfnamefont {J.}~\bibnamefont {Yang}},\
  }\href@noop {} {\bibfield  {journal} {\bibinfo  {journal} {The Astrophysical
  Journal Letters}\ }\textbf {\bibinfo {volume} {898}},\ \bibinfo {pages} {L19}
  (\bibinfo {year} {2020})}\BibitemShut {NoStop}%
\bibitem [{\citenamefont {Rushby}\ \emph {et~al.}(2020)\citenamefont {Rushby},
  \citenamefont {Shields}, \citenamefont {Wolf}, \citenamefont {Lagu{\"e}},\
  and\ \citenamefont {Burgasser}}]{rushby2020effect}%
  \BibitemOpen
  \bibfield  {author} {\bibinfo {author} {\bibfnamefont {A.~J.}\ \bibnamefont
  {Rushby}}, \bibinfo {author} {\bibfnamefont {A.~L.}\ \bibnamefont {Shields}},
  \bibinfo {author} {\bibfnamefont {E.~T.}\ \bibnamefont {Wolf}}, \bibinfo
  {author} {\bibfnamefont {M.}~\bibnamefont {Lagu{\"e}}}, \ and\ \bibinfo
  {author} {\bibfnamefont {A.}~\bibnamefont {Burgasser}},\ }\href@noop {}
  {\bibfield  {journal} {\bibinfo  {journal} {The Astrophysical Journal}\
  }\textbf {\bibinfo {volume} {904}},\ \bibinfo {pages} {124} (\bibinfo {year}
  {2020})}\BibitemShut {NoStop}%
\bibitem [{\citenamefont {North}(1975{\natexlab{a}})}]{north1975analytical}%
  \BibitemOpen
  \bibfield  {author} {\bibinfo {author} {\bibfnamefont {G.~R.}\ \bibnamefont
  {North}},\ }\href@noop {} {\bibfield  {journal} {\bibinfo  {journal} {Journal
  of the Atmospheric Sciences}\ }\textbf {\bibinfo {volume} {32}},\ \bibinfo
  {pages} {1301} (\bibinfo {year} {1975}{\natexlab{a}})}\BibitemShut {NoStop}%
\bibitem [{\citenamefont {North}(1984)}]{north1984}%
  \BibitemOpen
  \bibfield  {author} {\bibinfo {author} {\bibfnamefont {G.~R.}\ \bibnamefont
  {North}},\ }\href@noop {} {\bibfield  {journal} {\bibinfo  {journal} {Journal
  of the atmospheric sciences}\ }\textbf {\bibinfo {volume} {41}},\ \bibinfo
  {pages} {3390} (\bibinfo {year} {1984})}\BibitemShut {NoStop}%
\bibitem [{\citenamefont {Roe}\ and\ \citenamefont {Baker}(2010)}]{roe2010}%
  \BibitemOpen
  \bibfield  {author} {\bibinfo {author} {\bibfnamefont {G.~H.}\ \bibnamefont
  {Roe}}\ and\ \bibinfo {author} {\bibfnamefont {M.~B.}\ \bibnamefont
  {Baker}},\ }\href@noop {} {\bibfield  {journal} {\bibinfo  {journal} {Journal
  of climate}\ }\textbf {\bibinfo {volume} {23}},\ \bibinfo {pages} {4694}
  (\bibinfo {year} {2010})}\BibitemShut {NoStop}%
\bibitem [{\citenamefont {Walsh}(2017)}]{walsh2016diffusive}%
  \BibitemOpen
  \bibfield  {author} {\bibinfo {author} {\bibfnamefont {J.}~\bibnamefont
  {Walsh}},\ }\href@noop {} {\bibfield  {journal} {\bibinfo  {journal}
  {Discrete and Continuous Dynamical Systems B}\ }\textbf {\bibinfo {volume}
  {22}},\ \bibinfo {pages} {2687} (\bibinfo {year} {2017})}\BibitemShut
  {NoStop}%
\bibitem [{Note1()}]{Note1}%
  \BibitemOpen
  \bibinfo {note} {Note that annual average insolation distribution for rapidly
  rotating planets is symmetric about $90^\circ $ obliquity. The distribution
  when obliquity is $\beta $ is the same as when it is $180-\beta $. In the
  following article, we restrict our attention to obliquities between $0^\circ
  $ and $90^\circ $.}\BibitemShut {Stop}%
\bibitem [{\citenamefont {Nadeau}\ and\ \citenamefont
  {McGehee}(2017)}]{Nadeau2017}%
  \BibitemOpen
  \bibfield  {author} {\bibinfo {author} {\bibfnamefont {A.}~\bibnamefont
  {Nadeau}}\ and\ \bibinfo {author} {\bibfnamefont {R.}~\bibnamefont
  {McGehee}},\ }\href@noop {} {\bibfield  {journal} {\bibinfo  {journal}
  {Icarus}\ }\textbf {\bibinfo {volume} {291}},\ \bibinfo {pages} {46}
  (\bibinfo {year} {2017})}\BibitemShut {NoStop}%
\bibitem [{\citenamefont {Dobrovolskis}(2021)}]{dobrovolskis2020}%
  \BibitemOpen
  \bibfield  {author} {\bibinfo {author} {\bibfnamefont {A.~R.}\ \bibnamefont
  {Dobrovolskis}},\ }\href@noop {} {\bibfield  {journal} {\bibinfo  {journal}
  {Icarus}\ }\textbf {\bibinfo {volume} {363}},\ \bibinfo {pages} {114297}
  (\bibinfo {year} {2021})}\BibitemShut {NoStop}%
\bibitem [{\citenamefont {Budyko}(1969)}]{budyko1969effect}%
  \BibitemOpen
  \bibfield  {author} {\bibinfo {author} {\bibfnamefont {M.~I.}\ \bibnamefont
  {Budyko}},\ }\href@noop {} {\bibfield  {journal} {\bibinfo  {journal}
  {Tellus}\ }\textbf {\bibinfo {volume} {21}},\ \bibinfo {pages} {611}
  (\bibinfo {year} {1969})}\BibitemShut {NoStop}%
\bibitem [{\citenamefont {Sellers}(1969)}]{sellers1969global}%
  \BibitemOpen
  \bibfield  {author} {\bibinfo {author} {\bibfnamefont {W.~D.}\ \bibnamefont
  {Sellers}},\ }\href@noop {} {\bibfield  {journal} {\bibinfo  {journal}
  {Journal of Applied Meteorology}\ }\textbf {\bibinfo {volume} {8}},\ \bibinfo
  {pages} {392} (\bibinfo {year} {1969})}\BibitemShut {NoStop}%
\bibitem [{\citenamefont {North}(1975{\natexlab{b}})}]{north1975theory}%
  \BibitemOpen
  \bibfield  {author} {\bibinfo {author} {\bibfnamefont {G.~R.}\ \bibnamefont
  {North}},\ }\href@noop {} {\bibfield  {journal} {\bibinfo  {journal} {Journal
  of the Atmospheric Sciences}\ }\textbf {\bibinfo {volume} {32}},\ \bibinfo
  {pages} {2033} (\bibinfo {year} {1975}{\natexlab{b}})}\BibitemShut {NoStop}%
\bibitem [{\citenamefont {Held}\ and\ \citenamefont
  {Suarez}(1974)}]{held1974simple}%
  \BibitemOpen
  \bibfield  {author} {\bibinfo {author} {\bibfnamefont {I.~M.}\ \bibnamefont
  {Held}}\ and\ \bibinfo {author} {\bibfnamefont {M.~J.}\ \bibnamefont
  {Suarez}},\ }\href@noop {} {\bibfield  {journal} {\bibinfo  {journal}
  {Tellus}\ }\textbf {\bibinfo {volume} {26}},\ \bibinfo {pages} {613}
  (\bibinfo {year} {1974})}\BibitemShut {NoStop}%
\bibitem [{\citenamefont {Tung}(2007)}]{Tung2007}%
  \BibitemOpen
  \bibfield  {author} {\bibinfo {author} {\bibfnamefont {K.-K.}\ \bibnamefont
  {Tung}},\ }\href@noop {} {\emph {\bibinfo {title} {Topics in mathematical
  modeling}}}\ (\bibinfo  {publisher} {Princeton University Press Princeton,
  NJ},\ \bibinfo {year} {2007})\BibitemShut {NoStop}%
\bibitem [{\citenamefont {Widiasih}(2013)}]{widiasih2013dynamics}%
  \BibitemOpen
  \bibfield  {author} {\bibinfo {author} {\bibfnamefont {E.~R.}\ \bibnamefont
  {Widiasih}},\ }\href@noop {} {\bibfield  {journal} {\bibinfo  {journal} {SIAM
  Journal on Applied Dynamical Systems}\ }\textbf {\bibinfo {volume} {12}},\
  \bibinfo {pages} {2068} (\bibinfo {year} {2013})}\BibitemShut {NoStop}%
\bibitem [{\citenamefont {McGehee}\ and\ \citenamefont
  {Widiasih}(2014)}]{mcgehee2014}%
  \BibitemOpen
  \bibfield  {author} {\bibinfo {author} {\bibfnamefont {R.}~\bibnamefont
  {McGehee}}\ and\ \bibinfo {author} {\bibfnamefont {E.}~\bibnamefont
  {Widiasih}},\ }\href@noop {} {\bibfield  {journal} {\bibinfo  {journal} {SIAM
  Journal on Applied Dynamical Systems}\ }\textbf {\bibinfo {volume} {13}},\
  \bibinfo {pages} {518} (\bibinfo {year} {2014})}\BibitemShut {NoStop}%
\bibitem [{\citenamefont {Kaper}\ and\ \citenamefont
  {Engler}(2013)}]{kaper2013mathematics}%
  \BibitemOpen
  \bibfield  {author} {\bibinfo {author} {\bibfnamefont {H.}~\bibnamefont
  {Kaper}}\ and\ \bibinfo {author} {\bibfnamefont {H.}~\bibnamefont {Engler}},\
  }\href@noop {} {\emph {\bibinfo {title} {Mathematics and climate}}}\
  (\bibinfo  {publisher} {SIAM},\ \bibinfo {year} {2013})\BibitemShut {NoStop}%
\bibitem [{\citenamefont {Barry}\ \emph {et~al.}(2017)\citenamefont {Barry},
  \citenamefont {Widiasih},\ and\ \citenamefont
  {McGehee}}]{barry2014nonsmooth}%
  \BibitemOpen
  \bibfield  {author} {\bibinfo {author} {\bibfnamefont {A.~M.}\ \bibnamefont
  {Barry}}, \bibinfo {author} {\bibfnamefont {E.}~\bibnamefont {Widiasih}}, \
  and\ \bibinfo {author} {\bibfnamefont {R.}~\bibnamefont {McGehee}},\ }\href
  {\doibase 10.3934/dcdsb.2017125} {\bibfield  {journal} {\bibinfo  {journal}
  {Discrete and Continuous Dynamical Systems Series B}\ }\textbf {\bibinfo
  {volume} {22}},\ \bibinfo {pages} {2447} (\bibinfo {year}
  {2017})}\BibitemShut {NoStop}%
\bibitem [{\citenamefont {Cahalan}\ and\ \citenamefont
  {North}(1979)}]{cahalan1979stability}%
  \BibitemOpen
  \bibfield  {author} {\bibinfo {author} {\bibfnamefont {R.~F.}\ \bibnamefont
  {Cahalan}}\ and\ \bibinfo {author} {\bibfnamefont {G.~R.}\ \bibnamefont
  {North}},\ }\href@noop {} {\bibfield  {journal} {\bibinfo  {journal} {Journal
  of the Atmospheric Sciences}\ }\textbf {\bibinfo {volume} {36}},\ \bibinfo
  {pages} {1178} (\bibinfo {year} {1979})}\BibitemShut {NoStop}%
\bibitem [{\citenamefont {Walsh}\ and\ \citenamefont
  {Rackauckas}(2015)}]{walsh2015budyko}%
  \BibitemOpen
  \bibfield  {author} {\bibinfo {author} {\bibfnamefont {J.}~\bibnamefont
  {Walsh}}\ and\ \bibinfo {author} {\bibfnamefont {C.}~\bibnamefont
  {Rackauckas}},\ }\href@noop {} {\bibfield  {journal} {\bibinfo  {journal}
  {Discrete \& Continuous Dynamical Systems-B}\ }\textbf {\bibinfo {volume}
  {20}},\ \bibinfo {pages} {2187} (\bibinfo {year} {2015})}\BibitemShut
  {NoStop}%
\bibitem [{\citenamefont {Ward}(1974)}]{ward1974climatic}%
  \BibitemOpen
  \bibfield  {author} {\bibinfo {author} {\bibfnamefont {W.~R.}\ \bibnamefont
  {Ward}},\ }\href@noop {} {\bibfield  {journal} {\bibinfo  {journal} {Journal
  of Geophysical Research}\ }\textbf {\bibinfo {volume} {79}},\ \bibinfo
  {pages} {3375} (\bibinfo {year} {1974})}\BibitemShut {NoStop}%
\bibitem [{\citenamefont {Nadeau}\ and\ \citenamefont
  {McGehee}(2021)}]{Nadeau2021}%
  \BibitemOpen
  \bibfield  {author} {\bibinfo {author} {\bibfnamefont {A.}~\bibnamefont
  {Nadeau}}\ and\ \bibinfo {author} {\bibfnamefont {R.}~\bibnamefont
  {McGehee}},\ }\href@noop {} {\bibfield  {journal} {\bibinfo  {journal}
  {Journal of Mathematical Analysis and Applications}\ }\textbf {\bibinfo
  {volume} {To appear.}},\ \bibinfo {pages} {XX} (\bibinfo {year}
  {2021})}\BibitemShut {NoStop}%
\bibitem [{\citenamefont {Stone}(1978)}]{stone1978}%
  \BibitemOpen
  \bibfield  {author} {\bibinfo {author} {\bibfnamefont {P.~H.}\ \bibnamefont
  {Stone}},\ }\href@noop {} {\bibfield  {journal} {\bibinfo  {journal}
  {Dynamics of atmospheres and oceans}\ }\textbf {\bibinfo {volume} {2}},\
  \bibinfo {pages} {123} (\bibinfo {year} {1978})}\BibitemShut {NoStop}%
\bibitem [{\citenamefont {Wiens}(1999)}]{wiens1999log}%
  \BibitemOpen
  \bibfield  {author} {\bibinfo {author} {\bibfnamefont {B.~L.}\ \bibnamefont
  {Wiens}},\ }\href@noop {} {\bibfield  {journal} {\bibinfo  {journal} {The
  American Statistician}\ }\textbf {\bibinfo {volume} {53}},\ \bibinfo {pages}
  {89} (\bibinfo {year} {1999})}\BibitemShut {NoStop}%
\bibitem [{\citenamefont {Iaci}(2000)}]{iaci2000gamma}%
  \BibitemOpen
  \bibfield  {author} {\bibinfo {author} {\bibfnamefont {R.~J.}\ \bibnamefont
  {Iaci}},\ }\href@noop {} {\emph {\bibinfo {title} {The gamma distribution as
  an alternative to the lognormal distribution in environmental
  applications}}}\ (\bibinfo  {publisher} {University of Nevada, Las Vegas},\
  \bibinfo {year} {2000})\BibitemShut {NoStop}%
\bibitem [{\citenamefont {Strogatz}(2018)}]{strogatz2018nonlinear}%
  \BibitemOpen
  \bibfield  {author} {\bibinfo {author} {\bibfnamefont {S.~H.}\ \bibnamefont
  {Strogatz}},\ }\href@noop {} {\emph {\bibinfo {title} {Nonlinear dynamics and
  chaos with student solutions manual: With applications to physics, biology,
  chemistry, and engineering}}}\ (\bibinfo  {publisher} {CRC press},\ \bibinfo
  {year} {2018})\BibitemShut {NoStop}%
\bibitem [{\citenamefont {Ferreira}\ \emph {et~al.}(2014)\citenamefont
  {Ferreira}, \citenamefont {Marshall}, \citenamefont {O’Gorman},\ and\
  \citenamefont {Seager}}]{ferreira2014climate}%
  \BibitemOpen
  \bibfield  {author} {\bibinfo {author} {\bibfnamefont {D.}~\bibnamefont
  {Ferreira}}, \bibinfo {author} {\bibfnamefont {J.}~\bibnamefont {Marshall}},
  \bibinfo {author} {\bibfnamefont {P.~A.}\ \bibnamefont {O’Gorman}}, \ and\
  \bibinfo {author} {\bibfnamefont {S.}~\bibnamefont {Seager}},\ }\href@noop {}
  {\bibfield  {journal} {\bibinfo  {journal} {Icarus}\ }\textbf {\bibinfo
  {volume} {243}},\ \bibinfo {pages} {236} (\bibinfo {year}
  {2014})}\BibitemShut {NoStop}%
\bibitem [{\citenamefont {Wagner}\ and\ \citenamefont
  {Eisenman}(2015)}]{wagner2015climate}%
  \BibitemOpen
  \bibfield  {author} {\bibinfo {author} {\bibfnamefont {T.~J.}\ \bibnamefont
  {Wagner}}\ and\ \bibinfo {author} {\bibfnamefont {I.}~\bibnamefont
  {Eisenman}},\ }\href@noop {} {\bibfield  {journal} {\bibinfo  {journal}
  {Journal of Climate}\ }\textbf {\bibinfo {volume} {28}},\ \bibinfo {pages}
  {3998} (\bibinfo {year} {2015})}\BibitemShut {NoStop}%
\bibitem [{\citenamefont {Rose}\ and\ \citenamefont
  {Marshall}(2009)}]{rose2009ocean}%
  \BibitemOpen
  \bibfield  {author} {\bibinfo {author} {\bibfnamefont {B.~E.}\ \bibnamefont
  {Rose}}\ and\ \bibinfo {author} {\bibfnamefont {J.}~\bibnamefont
  {Marshall}},\ }\href@noop {} {\bibfield  {journal} {\bibinfo  {journal}
  {Journal of the atmospheric sciences}\ }\textbf {\bibinfo {volume} {66}},\
  \bibinfo {pages} {2828} (\bibinfo {year} {2009})}\BibitemShut {NoStop}%
\end{thebibliography}%

\section*{Appendix}

\label{sec:comparison}
In this section, we discuss a brief comparison of the analytical equilibrium solutions. As mentioned in Section \ref{sec:annual-avg-EB},
\cite{north1975analytical} comments that the divergence operator and the relaxation to the mean operator have the same effect on degree two polynomials when $\delta=\mu/6$. These operators are the same on the function subspace of degree two polynomials; however, solutions to the models live in larger spaces of functions, sometimes called solution spaces.  In particular, the equilibrium temperature solutions must be in the solution space so the relaxation model solution space contains piecewise polynomials of degree $2N$ and the diffusion model solution space contains bounded differentiable hypergeometric functions on the interval [0,1].  Degree two polynomials on [0,1] are a subset of both spaces. Even though solutions to the models are not restricted to the function subspace of degree two polynomials, the relation between $\delta$ and $\mu$ of $\delta=\mu/6$ remains a remarkably good approximation for comparing solutions between the models. We find that in the small and large limits of the heat transport efficiency parameter with other parameters fixed, the equilibrium temperature solutions from both models approach the same functions.

\begin{figure}
  \centering
  \includegraphics[width=0.95\linewidth]{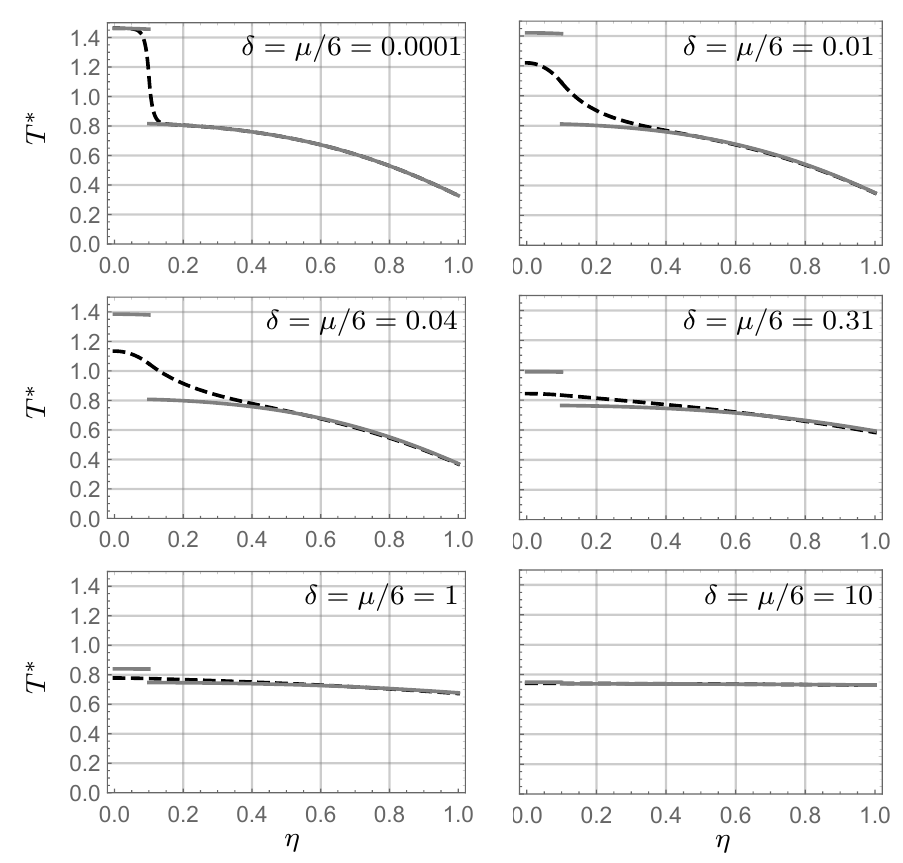}
\caption{{\small{Comparison of equilibrium temperature profiles $T^*_{\text{eq}}$ for the diffusion model (black dashed line) and the relaxation to the mean model (gray solid line). The plots are for Earth's albedo contrast and obliquity, $N=3$, ice line latitude $\eta=0.1$.}}} 
  \label{figure-appendix}
\end{figure}

As $\mu$ and $\delta$ approach zero, solutions of the relaxation to the mean model approach solutions to the diffusion model.  This can be seen directly by considering Equations \eqref{eq:nondim_model} and \eqref{eq:nondim_model_diff}.  As the parameters $\mu$ and $\delta$ approach zero, heat transport becomes negligible and the models become an energy balance equation where the latitude is simply a parameter. Convergence of the solutions across all latitudes as $\mu$ and $\delta$ approach 0 appears to be rather slow due to the continuous diffusive solution approaching the jump discontinuity in the relaxation to the mean solution at the ice line.

This behavior is illustrated in Figure \ref{figure-appendix}, wherein we compare the equilibrium temperature profile solutions of the two models with $N=3$ and Earth values for all parameters except $\delta$ and $\mu$.

As $\mu$ and $\delta$ become large, solutions of the relaxation to the mean model again approach solutions to the diffusion model. Intuitively this occurs because as the transport coefficients become large, heat is redistributed almost instantaneously, and the solutions approach the global mean temperature (see Equation \eqref{eq:Teq}).  Numerically we see that the solutions approach each other faster than they approach the global mean. Convergence for large $\mu$ and $\delta$ appears to be much faster compared to when the parameters approach zero.  For instance, for the parameters used in Figure \ref{figure-appendix}, the solutions are within 3\% of each other for all latitudes when $\delta=\mu/6=1$ and within 1\% of each other for all latitudes when $\delta=\mu/6=10$. Further mathematical investigation of these properties is a natural direction for future work.

\end{document}